\documentclass[10pt,aps,prb,superscriptaddress,floatfix,showpacs,longbibliography,notitlepage]{revtex4-1}
\usepackage{graphicx}
\usepackage{amsfonts}
\usepackage{amssymb}
\usepackage{amsmath}
\usepackage{txfonts}
\usepackage{lipsum}
\usepackage{color}
\usepackage{wasysym}
\usepackage{hyperref}
\usepackage{bbold}
\usepackage{etoolbox} 
\graphicspath{{figures/}}
\usepackage{feynmp-auto}
\usepackage{ulem}

\usepackage{mathtools}
\usepackage{multirow}
\usepackage{hhline}

\usepackage{tikz}

\usepackage{mathrsfs} 

\usepackage{letltxmacro}
\LetLtxMacro{\oldsqrt}{\sqrt}
\renewcommand{\sqrt}[2][\mkern8mu]{\mkern-6mu\mathop{}\oldsqrt[#1]{#2}}

\begin{document}
\title{Perturbative solution of fermionic sign problem in lattice Quantum Monte Carlo}

\author{Sergei Iskakov}
\affiliation{Department of Physics, University of Michigan, Ann Arbor, Michigan 48109, USA}

\author{Mikhail I. Katsnelson}
\affiliation{Institute for Molecules and Materials, Radboud University, 6525AJ Nijmegen, The Netherlands}

\author{Alexander I. Lichtenstein}
\email{alichten@physnet.uni-hamburg.de}
\affiliation{Institute of Theoretical Physics, University of Hamburg, 20355 Hamburg, Germany}
\affiliation{The Hamburg Centre for Ultrafast Imaging, Luruper Chaussee 149, 22761 Hamburg, Germany}

\begin{abstract}
We develop a strong-coupling perturbation scheme for a generic Hubbard model around a half-filled particle-hole-symmetric reference system,  which is free from the fermionic sign problem. The approach is based on the lattice determinantal Quantum Monte Carlo (QMC) method in continuous and discrete time versions for large periodic clusters in a fermionic bath. Considering the first-order perturbation in the shift of the chemical potential and of the second-neighbour hopping gives an accurate electronic spectral function for a parameter range corresponding to the optimally doped cuprate system for temperature of the order of $T=0.1t$, the region hardly accessible for the straightforward lattice QMC calculations. We discuss the formation of the pseudogap and the nodal-antinodal dichotomy for a doped Hubbard system in a strong-coupling regime with the interaction parameter $U$ equal to the bandwidth and the optimal value of the next-nearest-neighbor hopping parameter $t'$ for high-temperature superconducting cuprates.
\end{abstract}
\maketitle

\section*{INTRODUCTION}

Search for numerically exact solution of the $t-t'-U$ Hubbard model in thermodynamic limit at arbitrary interaction strength, long-range hoppings and doping $\delta$ or, equivalently, chemical potential $\mu$ at low temperature $T=1/\beta$ is tremendously difficult. Modern computational approaches, based on the lattice determinantal Quantum Monte Carlo (QMC) methods have seen incredible progress in the half-filled case without $t'$~\cite{SimonsPRX}, but 
face an unacceptable fermionic sign problem for a general case related to cuprate high-temperature superconductivity (HTSC) problem, which is the main factor restricting the accuracy of QMC calculations for interacting fermions\cite{sign1,sign2,Troyer_sign,Scalettar_sign2022}. 
A very important and largely unresolved problem is related to the next-nearest-neighbor hopping $t'$ in the Hubbard model and its role in the tendency towards superconductivity\cite{Deveraux_Tp,Deveraux_Hubb_Tp,White_Plaquette_Dw,Zhang_noSC,Harland16,Harland20,danilov2022}. 

On the other hand, the new class of diagrammatic Monte Carlo scheme\cite{Prokofev_sign_blessing} is claimed to have a ``sign blessing'' property which helps to reduce the effects of high-order diagrams.
The state-of-the-art diagrammatic Monte Carlo scheme in the connected determinant mode (C-DET)\cite{Rossi_PRL}   based on efficient Continuous Time Quantum Monte Carlo(CT-QMC) scheme in the weak coupling technique(CT-INT)~\cite{Rubtsov_CTINT} gives unprecedented accuracy for the doped Hubbard model\cite{Fedor_PG,Fedor_strong}. 
It becomes possible to study formation of the pseudogap already at the beginning of strong coupling case with $U/t=6$ \cite{Fedor_PG}.
Nevertheless, exponential convergence of the C-DET scheme for weak interactions \cite{Rossi_2017,Homotopic_QMC}, turns to a
divergence at large $U$ values due to poles in the complex $U$-plane\cite{Fedor_strong}. This means that calculations for interactions close to the bandwidth $U/t \approx 8$ and temperature
$T/t \approx 0.1$ are still within a prohibited area in the phase diagram\cite{Fedor_strong}.

There is a recent interesting attempt to use dynamical variational QMC scheme for the doped Hubbard model~\cite{Imada_VQMC2020,Tremblay_VQMC2022} which gives a very reasonable
description of the spectral function. The existence of the pseudogap can be explained in the simple model of electron fractionalization and appearing of ``dark'' fermion which is supported by $2\times 2$ cluster Dynamical Mean Field Theory (C-DMFT)~\cite{Civelli_CDMFT_HFermion2016,Harland16}. 
Moreover, the experimental RIXS spectrum~\cite{Huang_PG2022} of doped cuprate materials can be interpreted in such a theoretical model of the pseudogap formation.
The larger cluster in the C-DMFT scheme for the doped case has an unacceptable 
fermionic sign problem within the QMC scheme. Recently, the importance of vertex
corrections for pseudogap physics was discussed in the parquet formalism for dual fermions ~\cite{Krien_2022}.

\begin{figure}
\begin{center}
\includegraphics[width=0.9\textwidth]{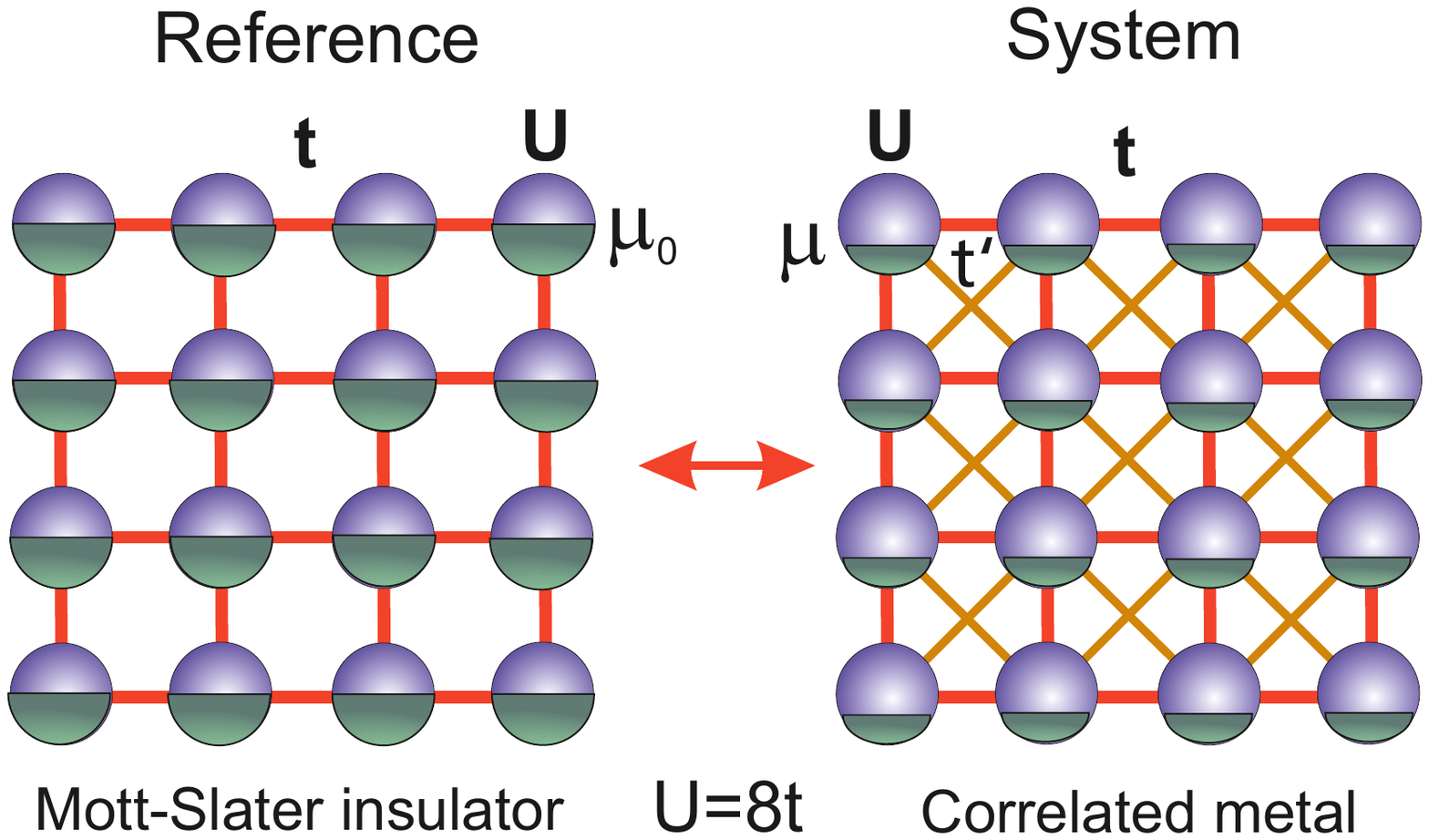}
\vskip -9cm
\includegraphics[width=0.4\textwidth]{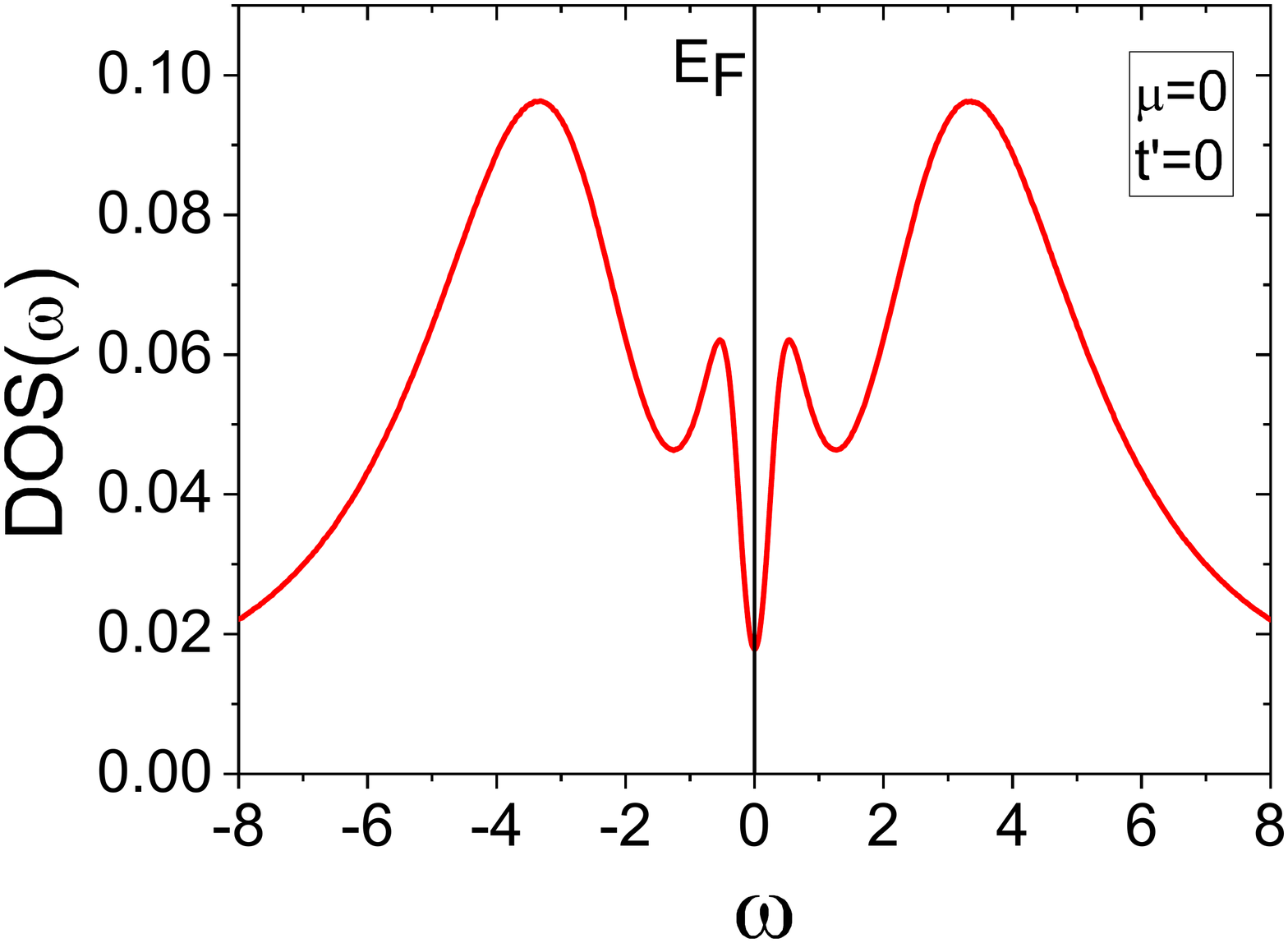}
\includegraphics[width=0.4\textwidth]{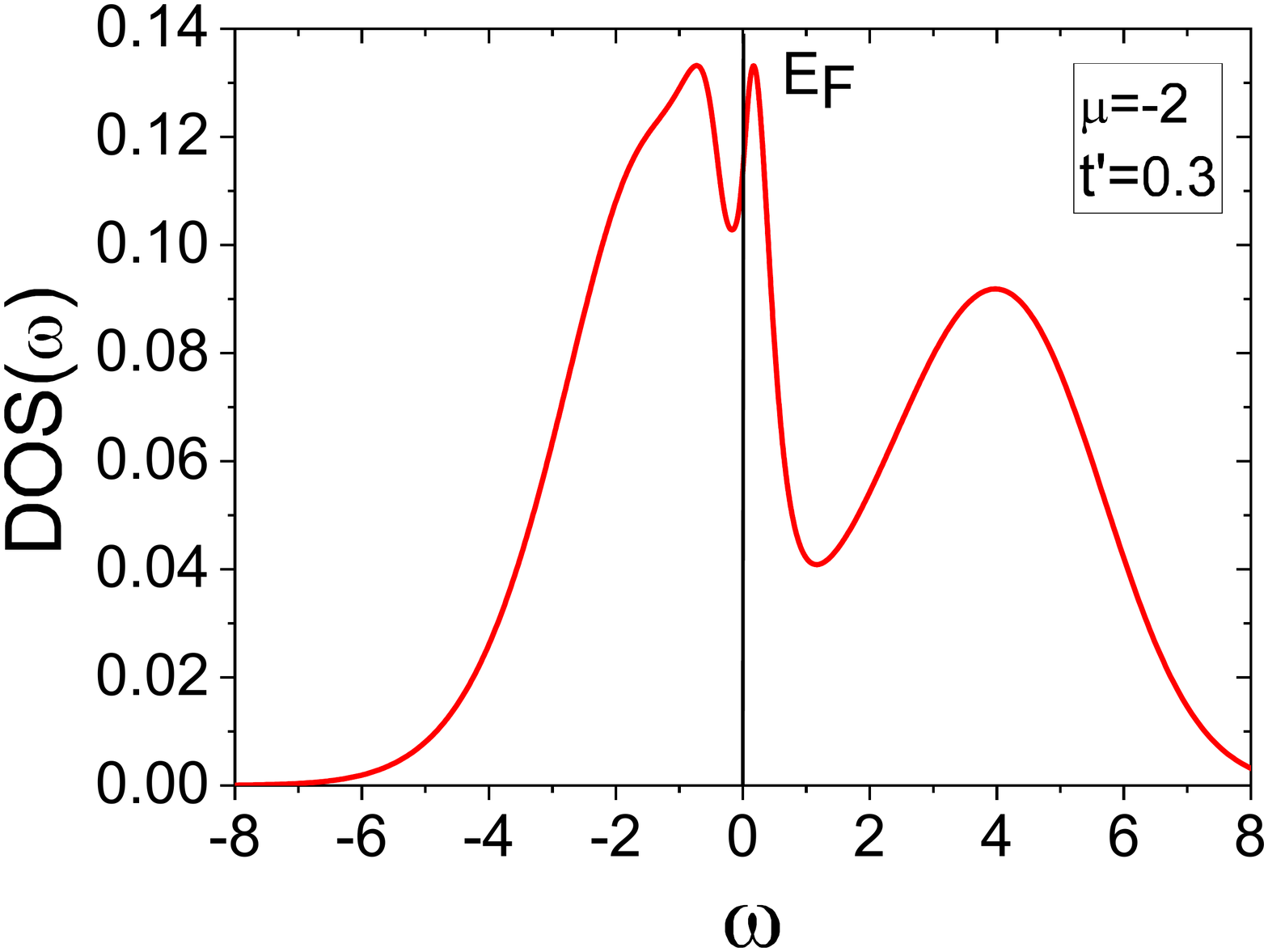}
\end{center}
 \caption{Schematic representation of a half-filled reference system for the doped square lattice. Bellow: calculated density of states (DOS) in presented scheme for $U=8$, Left: undoped case $\mu=0$ with $t'=0$, Right: doped case $\mu=-2$ with $t'/t=-0.3$}
\label{fig:refQMC}
\end{figure}

In this paper we discuss a different route to tackle the ``sign problem'' in the determinantal lattice QMC scheme and design a strong-coupling perturbative solution for a general Hubbard model.
The starting point is related to the ``reference system'' idea~\cite{Brener_refsys} which is basically quite simple and straightforward. The conventional choice of the noninteracting Hamiltonian as the reference system for the perturbation \cite{AGD} is justified by Wick's theorem which allows to calculate exactly any many-particle Green's functions: they are all expressed in terms of single-particle Green's functions. The choice of single-site approximation like dynamical mean-field theory \cite{RevModPhys.68.13} as the reference system leads to the dual fermion technique \cite{RKL08,Brener_refsys}. Actually, the reference system can be arbitrary assuming that we can calculate its Green's functions of arbitrary order. Of course, in practice this is hardly doable. 

At the same time, sometimes even taking into account the simplest, first-order diagram, seems to be quite successful. In the conventional weak-coupling expansion it is equivalent to the famous Hartree-Fock approximation \cite{Bethe} which is able to catch a lot of important many-body physics including e.g. superconductivity within the BCS model. It can be shown \cite{IK1,IK2,IK3} that the unrestricted Hartree-Fock trial wave function is optimal within a very broad class of variational ground-state wave functions for different physical systems. 
It is worthwhile to mention here the very successful Peierls-Feynman-Bogoliubov variational principle~\cite{Peierls_1938,Feynman_1972,Bogolyubov_1958} which can be formulated in the path-integral scheme. In this case, a good variational estimate of the system's free energy $F$ with the Hamiltonian $H_1$ is achieved on an optimal reference system with the Hamiltonian $H_0$, namely 
$F_1 \leq F_0+\langle H_1-H_0 \rangle _0$.
One can hope therefore that even first-order corrections to the properly chosen reference system will already give a rich and adequate enough physical picture. At least, this is definitely an assumption worth to be checked. 

Here our reference system corresponds to the half-filled  ($\mu=0$) particle-hole symmetric ($t'=0$) case  (Figure \ref{fig:refQMC}) where lattice Monte Carlo has no sign problem and the numerically exact solution for any practical value of $U$ is possible within a broad range of temperatures~\cite{DQMC_Scalettar}. Then we apply the lattice dual fermion perturbation theory \cite{RKL08,DF_Rev,Brener_refsys} to find the first-order perturbative corrections in $\mu$ and $t'$. To this aim, it is sufficient to calculate the two-particle Green's function, or, equivalently, four-leg vertex, which can be done accurately enough within the continuous time Quantum Monte Carlo. This approach can be considered as a far-going generalization of the Hartree-Fock approximation to the case of a dynamical effective interaction. Our reference system already has the main correlation effects in the lattice and shows characteristic ``four-peak'' structure with high-energy Hubbard bands around $\pm U/2$ and antiferromagnetic Slater bands close to the insulating gap
(which can be seen in the density of states in Figure \ref{fig:refQMC}, left panel).
After the dual fermion perturbation scheme the correlated metallic states
with the DMFT-like "three peak" structure appear with a pseudogap-like feature at a high temperature (the density of states in Figure \ref{fig:refQMC}, right panel).
The results for the strong-coupling case ($U=W=8t$) with practically interesting values of the chemical potential and next-nearest-neighbour hoppings corresponding to cuprate superconductors have shown formation of a pseudogap and nodal-antinodal dichotomy (that is, well-defined quasiparticles in the nodal part of the Fermi surface and strong quasiparticle damping for the antinodal part) which makes this approximation a perspective for practical applications. 

\section*{RESULTS}

We have calculated the Green's function for the doped two-dimensional Hubbard model for a periodic $8\times8$ system with $U/t=8$, $t'/t=-0.3$ and $\mu=-2.0$ in units of $t$ for $\beta=10$ (in units $1/t$) using a CT-INT version of the CT-QMC scheme~\cite{Rubtsov_CTINT}. Note that for the non-interacting Green's function we used the infinite-lattice limit with periodic boundary conditions
for the calculated $8\times8$ system (see Section METHODS). 
This scheme reduces the cluster-size dependence for the bare Green's function: in particular, the local one does not depend at all on the choice of the ``simulation box''. On the other hand, it may underestimate the effect of $U$-interactions, since it appears only in the calculated cluster. This may explain a small gap in the half-filled
reference system compare to a standard lattice determinantal QMC scheme~\cite{Rost_QMC}.

\begin{figure}
\begin{center}
\includegraphics[width=0.8\linewidth]{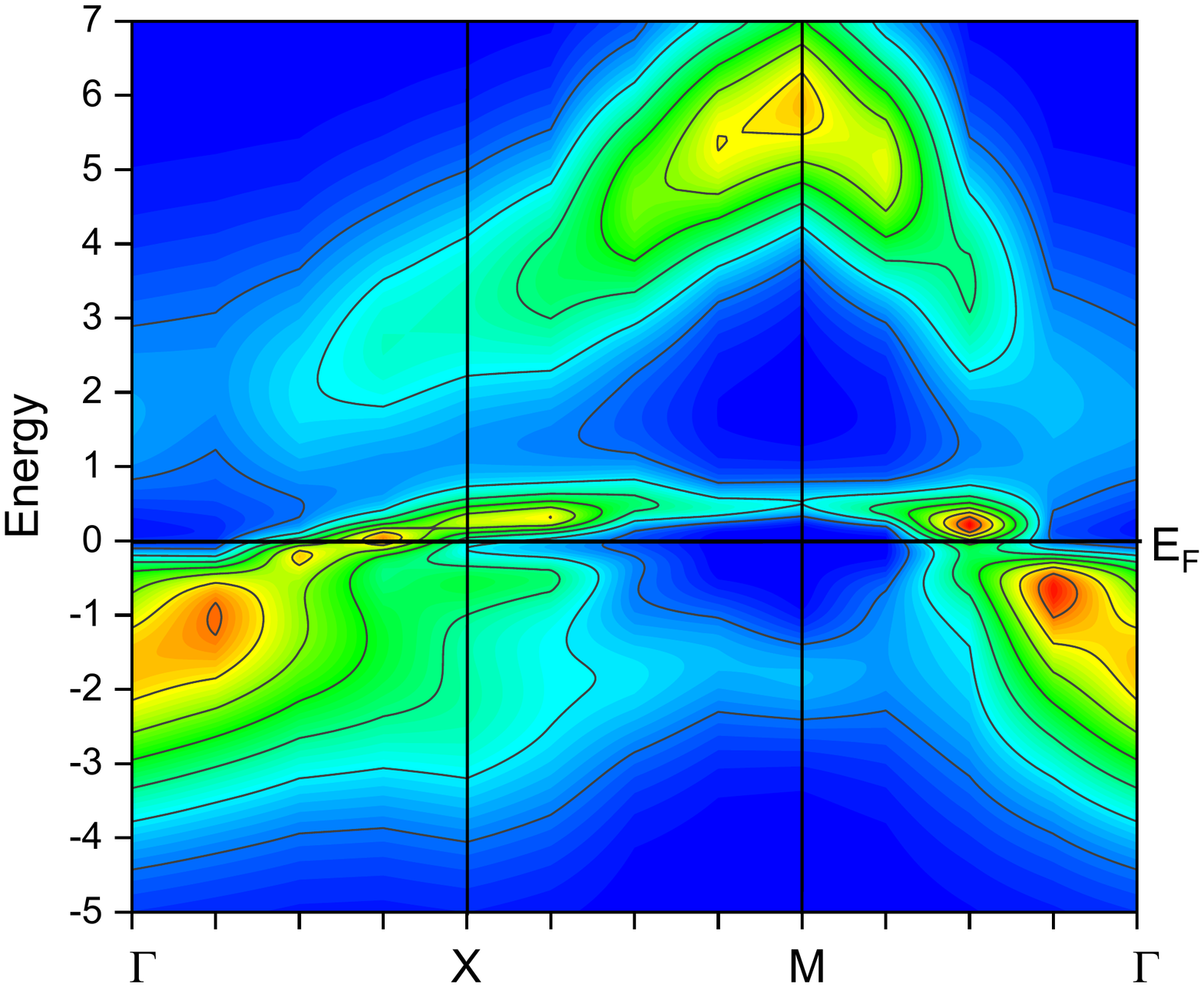}
\end{center}
\caption{Spectral function $-1/\pi \Im G({\bold k},\omega)$ for dual fermion QMC (CT-INT)  for $(8\times8)$ lattice with $U=8$, $t'/t=-0.3$, $\mu=-2.0$, and
$\beta=10$.}
\label{fig:Gk8x8}
\end{figure}

The results for the first-order dual-fermion perturbation from the half-filled system indicate
the formation of correlated pseudogap electronic structure. Figure \ref{fig:Gk8x8} shows the color map of the spectral function along the irreducible path ($\Gamma-X-M-\Gamma$) in the square Brillouin zone. For analytical continuation the newly developed scheme~\cite{Gull_MaxEnt} was used.

Several characteristic features of the correlated metallic phase in generic cuprate systems can be detected: formation of an extended pseudogap region around the $X$-point towards the $M$-point, a shadow antiferromagnetic band at energy $-2 t$ near the $M$-point, a strongly renormalized metallic band near the nodal point around $\Gamma M/2$. Overall, the spectral function for $U=W$ clearly shows strong correlation features of the electronic structure far beyond a simple renormalized-band paradigm.

\begin{figure}
\begin{center}
\includegraphics[width=0.6\linewidth]{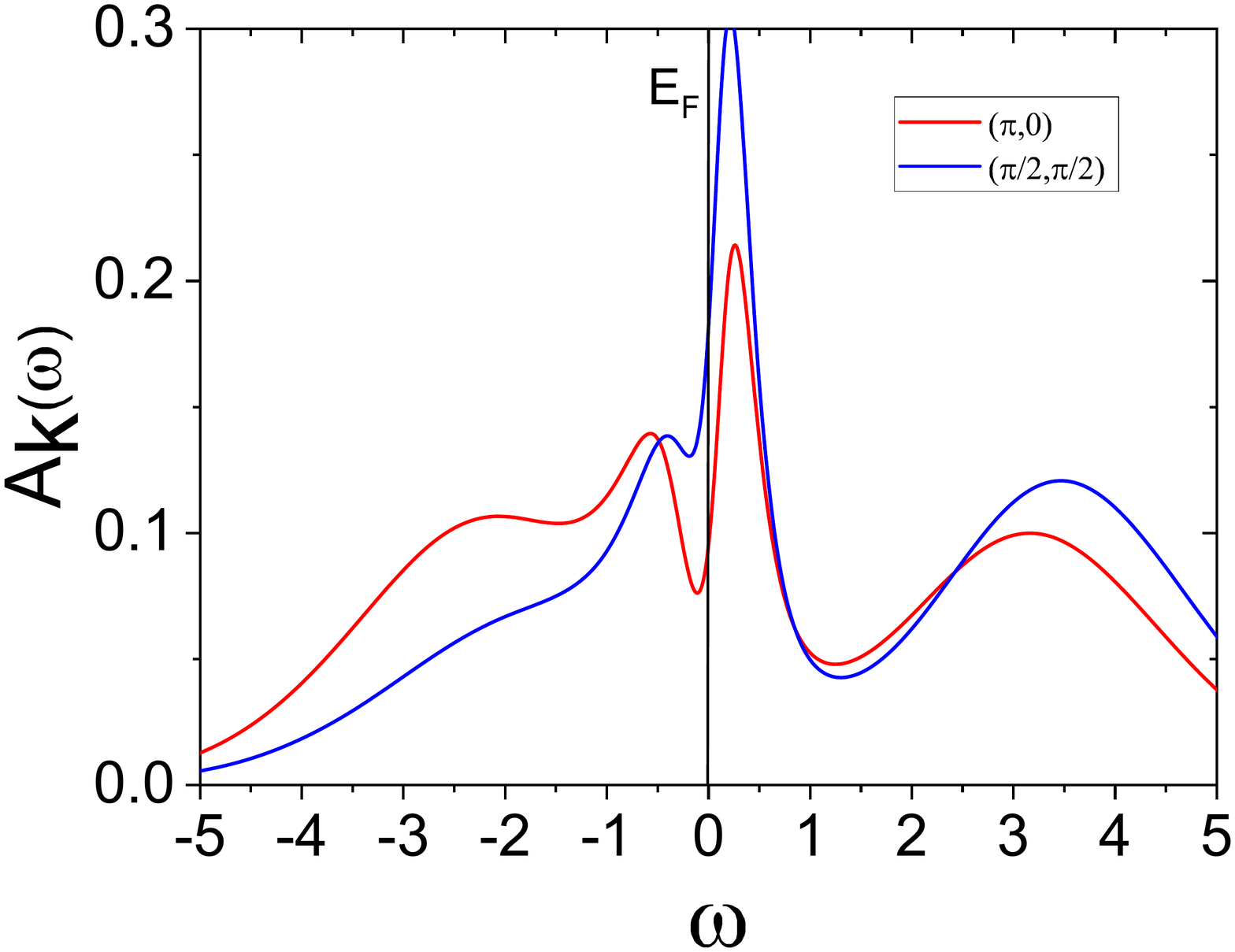}
\end{center}
\caption{Spectral function $-1/\pi \Im G({\bold k},\omega)$ for two different k-points  corresponds to anti-nodal and nodal k-points dual fermion QMC (CT-INT)  for $(8\times8)$ lattice with $U/t=8$ $t'/t=-0.3$, $\mu=-2.0$ and $\beta=10$ .}
\label{fig:Gk8x8NAN}
\end{figure}

\begin{figure}
\begin{center}
\includegraphics[width=0.33\linewidth]{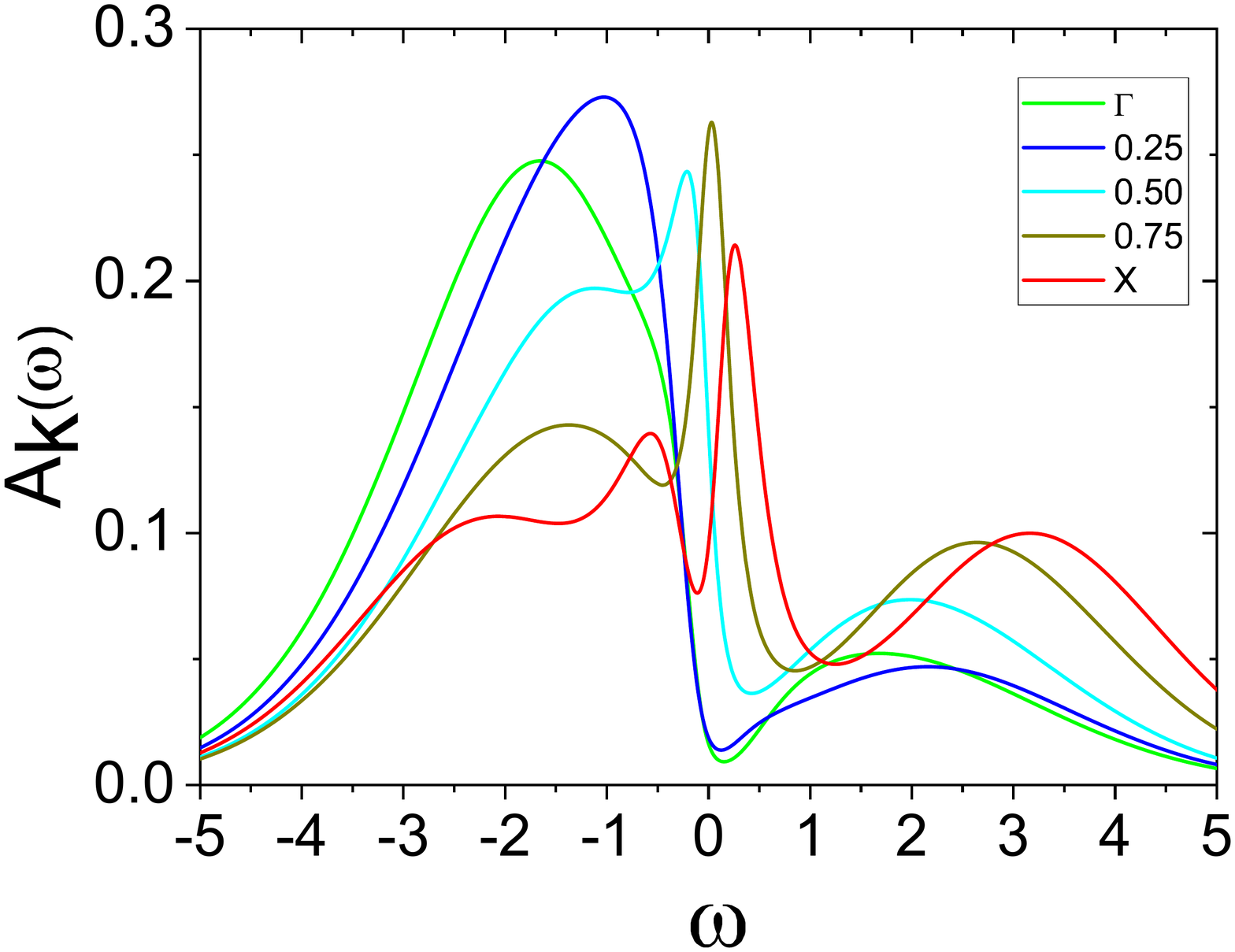}
\includegraphics[width=0.33\linewidth]{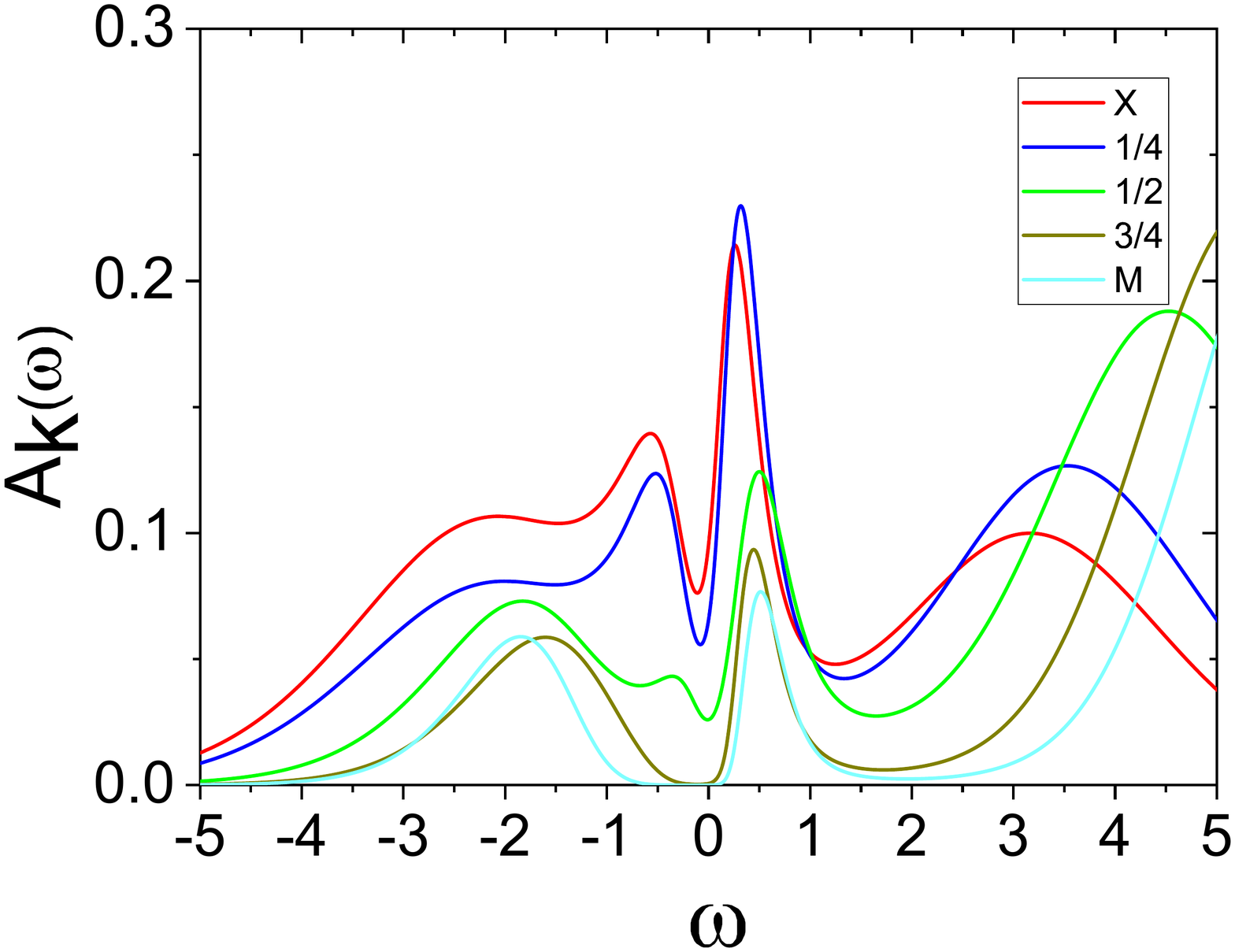}
\includegraphics[width=0.33\linewidth]{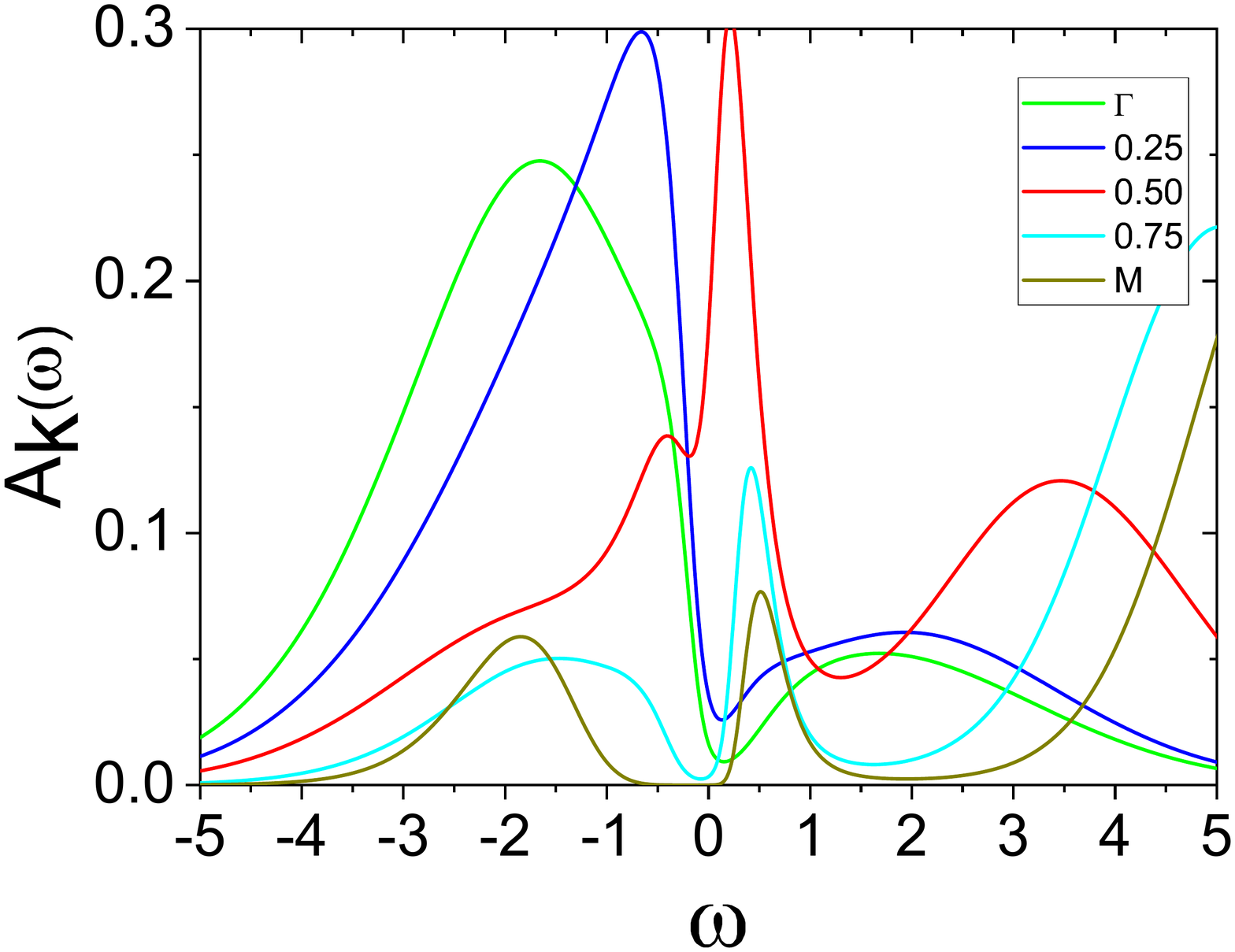}
\end{center}
\caption{Spectral function $-1/\pi \Im G({\bold k},\omega)$ for three different k-directions in the Brillouin Zone, (left) $\Gamma-X$, (middle) $X-M$ and (right) ($\Gamma-M$) dual fermion QMC (CT-INT)  for $(8\times8)$ lattice with $U/t=8$ $t'/t=-0.3$, $\mu=-2.0$ and $\beta=10$ .}
\label{fig:Gk8x8BZ}
\end{figure}

\begin{figure}
\begin{center}
\includegraphics[width=0.45\linewidth]{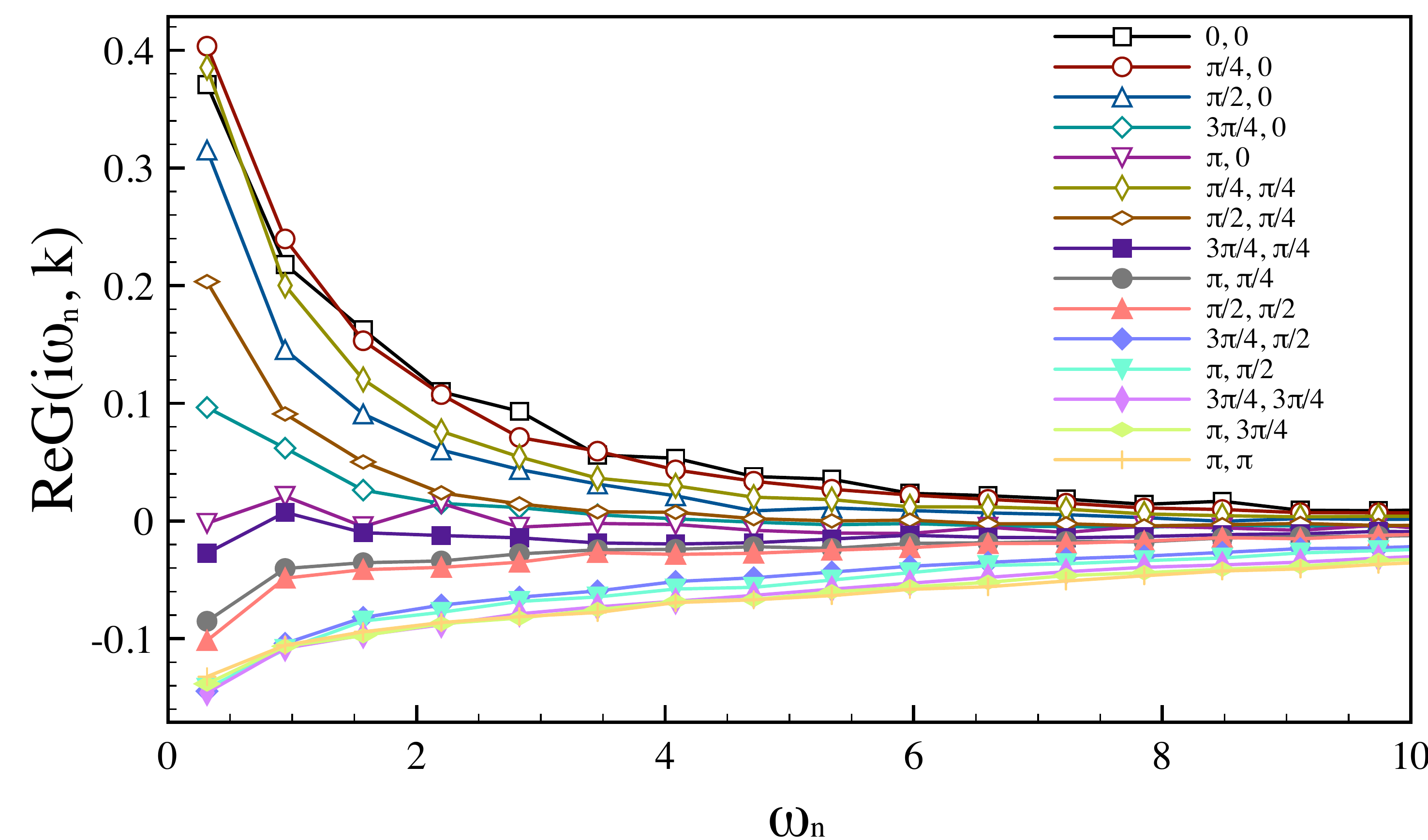}
\includegraphics[width=0.45\linewidth]{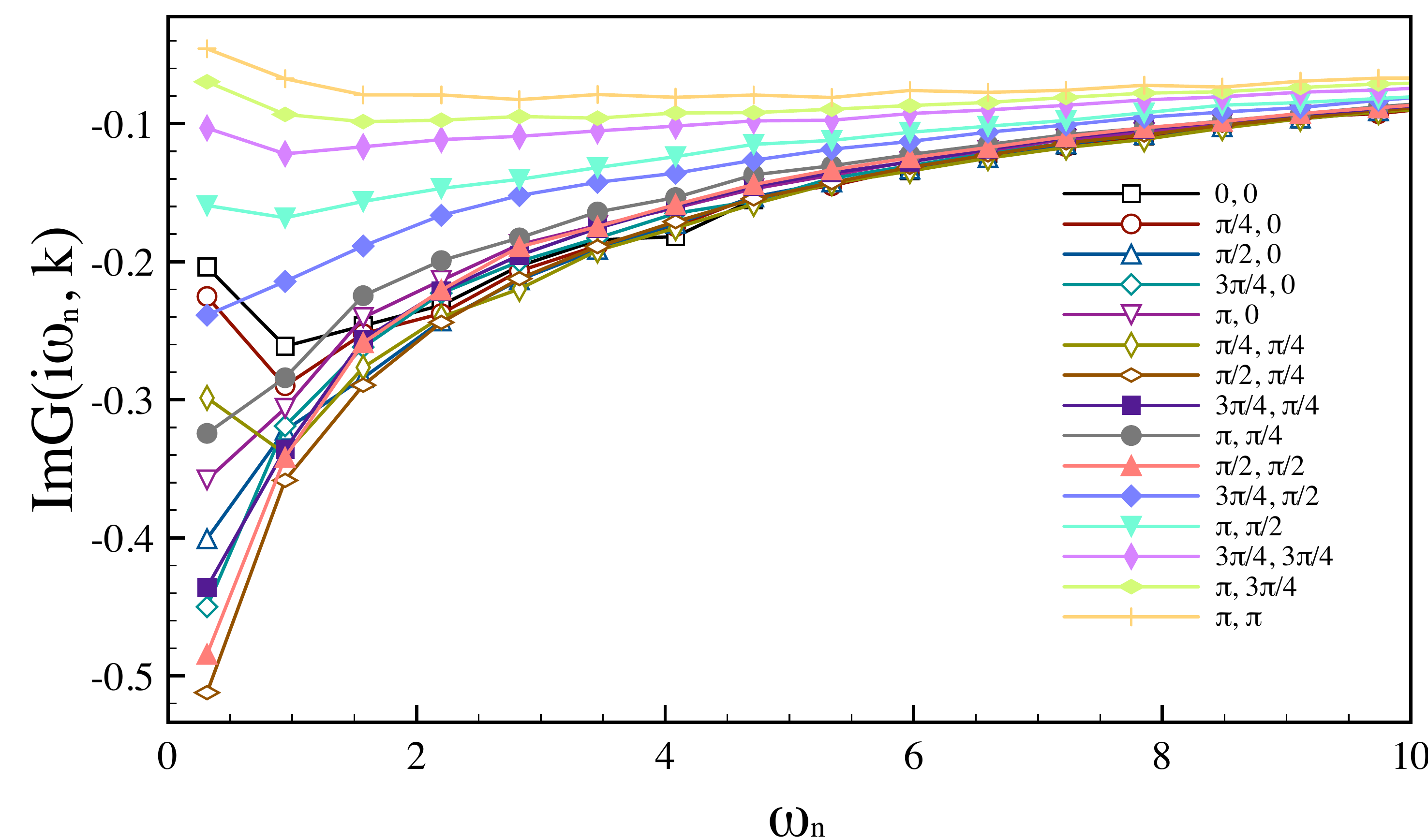}
\end{center}
\caption{Green's function $G({\bold k},\omega_n)$ on the Matsubara axes for all 16 non-equivalent k-points in the Brillouin Zone for  $8\times8$ system, (left) Real part and (right) imaginary part for dual fermion QMC (CT-INT) with $U/t=8$ $t'/t=-0.3$, $\mu=-2.0$ and $\beta=10$ .}
\label{fig:Gk8x8Matsub}
\end{figure}

In order to see more clearly the pseudogap and nodal-antinodal dichotomy we plot the energy dependence of two spectral functions at the $X$- and $\Gamma M/2$-points in the Brillouin zone (Figure \ref{fig:Gk8x8NAN}). While at the $X=(\pi,0)$-point there is a reasonably deep 
pseudogap fromation already at $\beta=10$, the nodal spectral function at $(\pi/2,\pi/2)$ has correlated metallic behaviour. 
A more unusual feature of the strong-coupling spectral function in Figure \ref{fig:Gk8x8} is related with a "shark mouth" pseudogap dip starting 
at $X$ in the direction to $M$ until the half way. One can see from the energy dependence of the spectral function in the direction of $X-M$
(Figure \ref{fig:Gk8x8BZ}, middle panel) that the pseudogap splitting of the sharp quasiparticle peak at zero for the $XM/4$ point is even larger than at the $X$-point.
The same feature was observed for a self-energy in the diagrammatic Monte Carlo (C-DET) investigation of the doped Hubbard model at $U/t=6$ \cite{Fedor_PG}.
We would like to point out that all these effects are not simply an artifact of the analytical continuation with the MaxEnt scheme and
can be detected by inspection of the original complex Matsubara Green's function from DF-QMC calculations (Figure \ref{fig:Gk8x8Matsub}).
If we compare the $X=(\pi,0)$ and $XM/4=(\pi,\pi/4)$ points then both quasiparticle peaks located almost at the Fermi energy (the real part of 
$G({\bold k},\omega_n)$ is close to zero, but the pseudogap or upturn of the imaginary part of $G({\mathbf{k}},\omega_n)$ for the first Matsubara frequencies are more pronounced at the $(\pi,\pi/4)$-point. We have also checked this characteristic feature for the Hirsch-Fye QMC scheme~\cite{Hirsch-Fye} and different MaxEnt implementations. The general structure of this spectral
function is similar to recent results of dynamical variational Monte Carlo schemes~\cite{Imada_VQMC2020,Tremblay_VQMC2022}. 

\begin{figure}
\begin{center}
\includegraphics[width=0.48\linewidth]{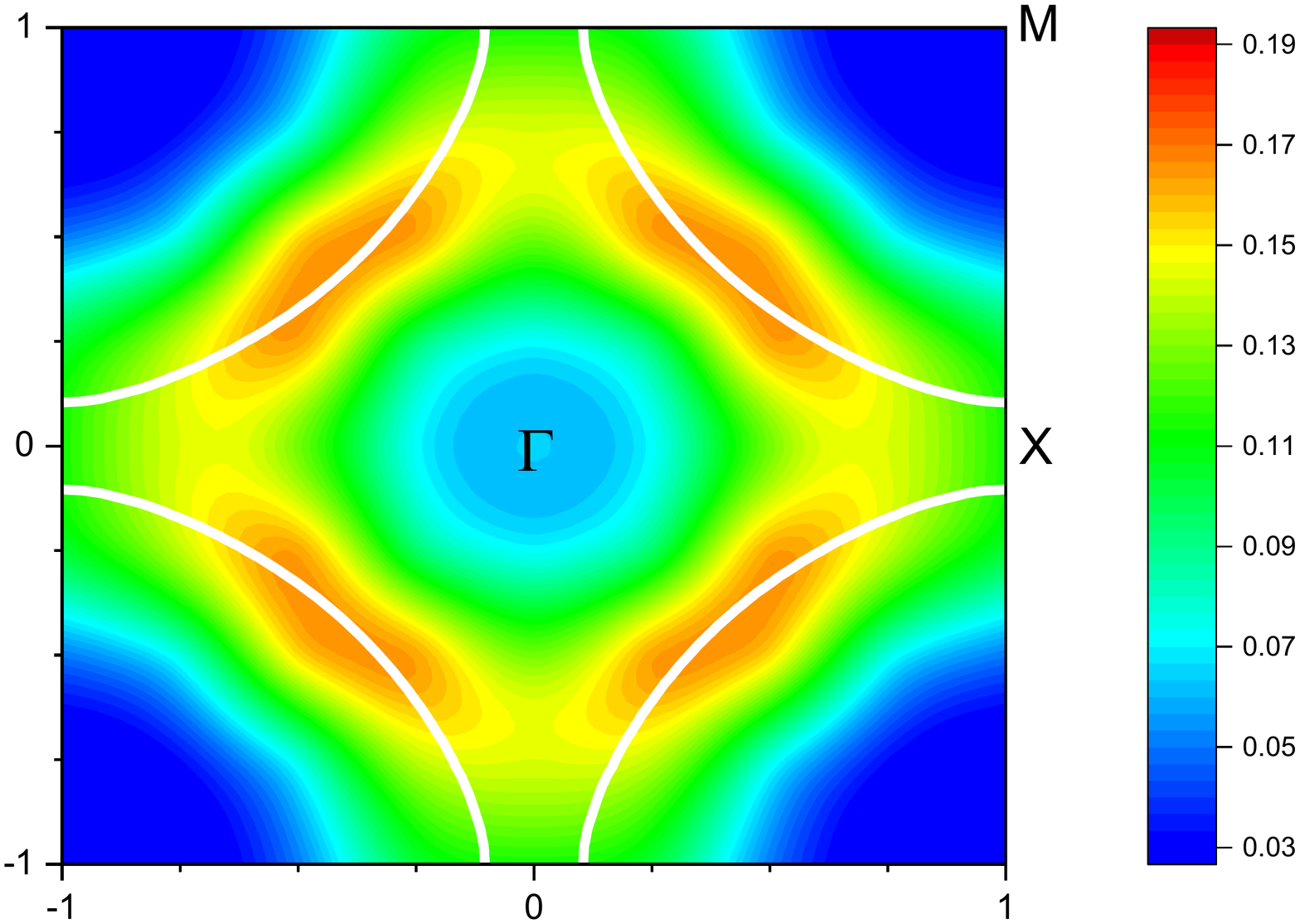}
\includegraphics[width=0.48\linewidth]{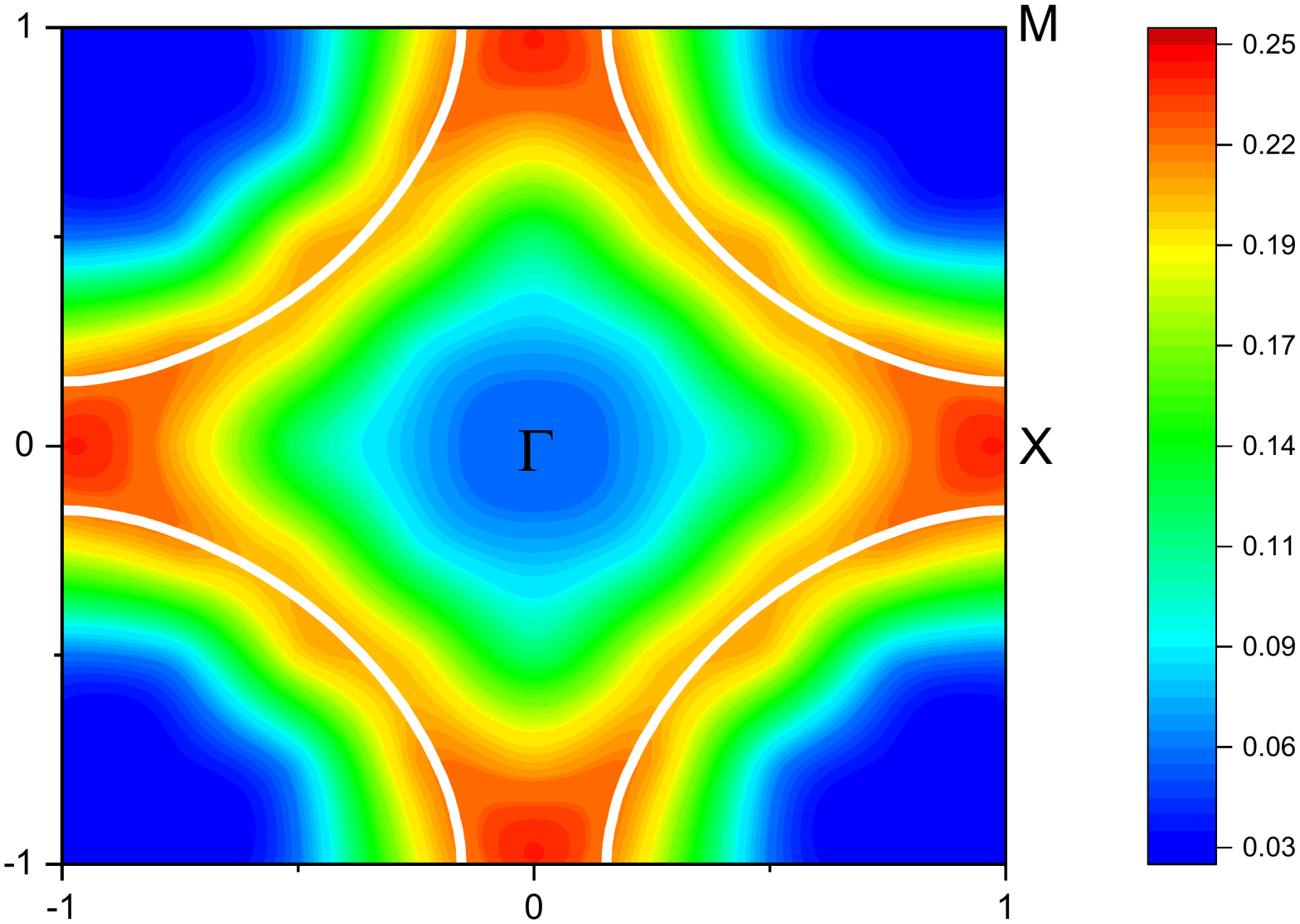}
\end{center}
\caption{Spectral function of the square-lattice Hubbard model as a function of momentum at the first Matsubara frequency $G({\bold k},\omega_0)$  for dual fermion QMC with $t'/t=-0.3$, $\beta=10$ and  $U/t=8$, $\mu=-2.0$  (left) $U/t=5.6$, $\mu=-0.9$  (right) . The non-interacting Fermi surface with the same doping 
is shown for comparisson as a white contour.}
\label{fig:Gk8x8FS}
\end{figure}

We plot a broadened Fermi surface using the momentum-dependent spectral function for the first Matsubara frequency (Figure \ref{fig:Gk8x8FS}). Comparison with the non-interacting tight-binding Fermi surface for the same doping shows a large region of the pseudogap around the $X$-point and formation of Fermi arcs near the nodal point. Moreover, one can understand that the pseudogap is more pronounced a bit away from the $X$-point towards the $M$-point, where the non-interacting Fermi surface
crosses the Brillouin zone. We also compare the Fermi surface plot for smaller values of $U/t=5.6$,
which was investigated in the diagrammatic Monte Carlo technique~\cite{Wei_point,Rossi_2020}; this value is related to a plaquette degenerate point~\cite{danilov2022}. While the Fermi surface for small $U/t=5.6$ agrees well with the results of the diagrammatic Monte Carlo approach~\cite{Rossi_2020} and resembles the tight-binding one with only large broadening around the $X$-point, the $U/t=8$ results show already a formation of the pseudogap and Fermi arcs, that is, a nodal-antinodal dichotomy.

\subsection*{}

\section*{DISCUSSION}
We developed, for Hubbard-like correlated lattice models, the first-order strong-coupling dual fermion expansion in the shift of the chemical potential (doping) and in the second-neighbour hoppings ($t'$). The starting reference point corresponds to the half-filled particle-hole symmetric system which can be calculated numerically exactly, without the fermionic sign problem. For physically interesting parameter range of cuprate-like systems (around 10$\%$ doping and $t'/t=-0.3$0 we can obtain a reasonable Green's function for a periodic $8\times8$ lattice for the temperature $T=0.1t$ .
The formation of the pseudogap around the antinodal $X$-point and the nodal-antinodal dichotomy are clearly seen in the present approach.

We would like to point out a few main reasons why such a ``super-perturbation'' scheme works: first of all, the reference system already
contains the main correlation effects which result in the four-peak structure of the density of states for the half-filled lattice Monte-Carlo calculations~\cite{Rost_QMC}; second, the first-order strong-coupling perturbation relies on the lattice four-point vertex 
$\gamma_{1 2 3 4}$ (Eq. (\ref{eqn::chiph})) which is obtained numerically exactly and has all the information about the spin and charge
susceptibilities of the lattice; and third, if the dual perturbation Green's function $\tilde{G}^0_{12}$ (Eq. (\ref{eqn::df::Gdbare})) is relatively small, results will be reasonable. The complicated question of convergence for such a dual-fermion 
perturbation can be checked numerically by calculating the second-order contribution in $\Tilde{\Sigma}_{12}$. For this term one needs to calculate in the lattice QMC a six-point vertex $\gamma^{(6)}$ which will be also a direction of future developments.
In principle, one can also discuss an instability towards antiferromagnetism or d-wave superconductivity, introducing symmetry-breaking fields~\cite{White_Plaquette_Dw}, which we also plan to investigate. 

It is worthwhile to mention that for the starting reference system we can choose not only the half-filled case, but any doped case where the sign
problem is mild, so we can use a QMC calculation to expand this numerically exact solution to "Terra incognita" regions where
the sign problem is unacceptable for direct QMC calculations.

\section*{METHODS}

We start with the general version of the cluster dual fermion scheme~\cite{RKL08,HH_CDF} for $t-t'-U$ square lattice Hubbard model.
The general strategy of the dual fermion approach as a strong coupling theory is related to formally exact expansion around arbitrary reference system~\cite{Brener_refsys}

\subsection*{Hamiltonian}
The simplest model describing interacting fermions on a
lattice is the single band Hubbard model, defined by the Hamiltonian
\begin{eqnarray}
{\hat H_\alpha}=-\sum_{i,j,\sigma} 
 t^\alpha_{ij}c_{{i}\sigma}^{\dagger}c_{{j}\sigma}^{\phantom{\dagger}} 
             + \sum_{ i} U  (n_{{i}\uparrow } - \frac{1}{2}  )    (n_{{i}\downarrow } - \frac{1}{2}   ) 
       \label{Hub}
\end{eqnarray} 
where $t_{ij}$ is the hopping matrix elements including
the chemical potential $\mu$ in the diagonal elements.
\begin{equation}
t^\alpha_{ij}=
\begin{cases}
    t & \text{if $i$ and $j$ are nearest neighbours},\\
    \alpha t' & \text{if $i$ and $j$ are next nearest neighbours},\\
   \alpha \mu & \text{if $i$ = $j$} ,\\
      0 & \text{otherwise},
\end{cases}
\label{tij}
\end{equation}
where $n_{{i}\sigma }= c_{{i}\sigma}^{\dagger} c_{{i}\sigma}^{\phantom{\dagger}} $. We introduce a "scailing" parameter $\alpha=0,1$, which defined a reference system $H_0$ for  $\alpha=0$ which corresponds to the half-filled Hubbard model ($\mu_0=0$) with only nearest neighbours hoppings ($t'_0=0$) and final system $H_1$ for  $\alpha=1$ for given
$\mu$ and $t'$. Notes, that long-range hoping parameters can be trivially included similar to $t'$.

\subsection*{Real space scheme}
For the super-perturbation in the lattice Monte-Carlo scheme we use a general dual-fermion expansion around arbitrary reference system within
the path-integral formalism ~\cite{RKL08,Brener_refsys} similar to a strong coupling expansion  ~\cite{Pairault_PRL}. 
In this case our $N \times N$ lattice and corresponding reference systems
represent $N \times N$-part which we cut from infinite lattice and periodise the bare Green's function $\mathcal{G}_\alpha $.
The general lattice action for discretise $N \times N \times L$ space-time lattice (for CT-INT scheme imaginary time space $\tau$ is continuous in the $[0,\beta )$ interval) with Hamiltonian  Eq.~(\ref{Hub}) reads  
\begin{equation}
S_\alpha [c^{ \ast } ,c] = -\sum _{1,2} c_{ 1}^{ \ast } \; (\mathcal{G}_{\alpha})_{12}^{ -1}\; c_{ 2} +\frac{1}{4} \sum _{1234  } U_{1 2 3 4} c_{1}^{ \ast } c_{2}^{ \ast } c_{4} c_{3}\, .
\label{eqn:lat_action}
\end{equation}
 In order to keep the notation simple, it is useful to introduce the combined index $\left \vert 1\right \rangle  \equiv \left \vert i ,\tau   ,\sigma   \right \rangle $ ($i$ being the site index suppressed above) while assuming summation over repeated indices. 
 
To calculate the bare propagators $(\mathcal{G}_{\alpha})_{12}$ we start from the $N\times N$ cluster which is cut from infinite lattice and then force
translation symmetry and periodic boundary condition on the  finite $N\times N$ system. This procedure is easy to realize in the k-space, by
doing first a double Fourier transform of the bare Green's function for non-periodic $N\times N$ cluster 
$\mathcal{G}^{\alpha}_{\mathbf{k},\mathbf{k'}}$ and then keep only periodic part, $\mathcal{G}^{\alpha}_{\mathbf{k}}\delta_{\mathbf{k},\mathbf{k'}}$.

Perturbation matrix of one-electron part of Action:
\begin{equation}
\tilde {t} =  \mathcal{G}_{0}^{ -1} -\mathcal{G}_{1}^{ -1} \, .
\label{tpertG}
\end{equation}

The dual action in paramagnetic state reads
\begin{equation}
\tilde{S} [d^{ \ast } ,d] = -\sum _{ \mathbf{12}\, \nu \sigma }d_{1 \nu \sigma }^{ \ast } \; (\tilde{G}^0_{\nu})^{-1}_{12  }\; d_{2 \nu \sigma } +\frac{1}{4} \sum _{1234  } \gamma_{1 2 3 4} d_{1}^{ \ast } d_{2}^{ \ast } d_{3} d_{4},
\label{eqn::df::dual_action}
\end{equation}
where the bare dual Green's function has the following matrix form:
\begin{equation}
\tilde{G}^0_{12 }  =\left [\,\tilde{t}^{ -1} -\hat{g}\,\right]^{ -1}_{12}
\label{eqn::df::Gdbare}
\end{equation}
with $g$ being exact Green's matrix of the interacting referemnce system.

We used the following notation for the four-point vertex:
\begin{align}
\gamma_{1 2 3 4}  = \langle c_{1} c_{2}^{ \ast } c_{4} c_{3}^{ \ast } \rangle -
 \langle c_{1}  c_{2}^{ \ast } \rangle  \langle      c_{4}   c_{3}^{ \ast }\rangle  +
  \langle c_{1} c_{3}^{ \ast }    \rangle \langle      c_{4}   c_{2}^{ \ast } \rangle 
\label{eqn::chiph}
\end{align}

\begin{figure}
 \centering
 \includegraphics[width=0.15\textwidth,angle=0]{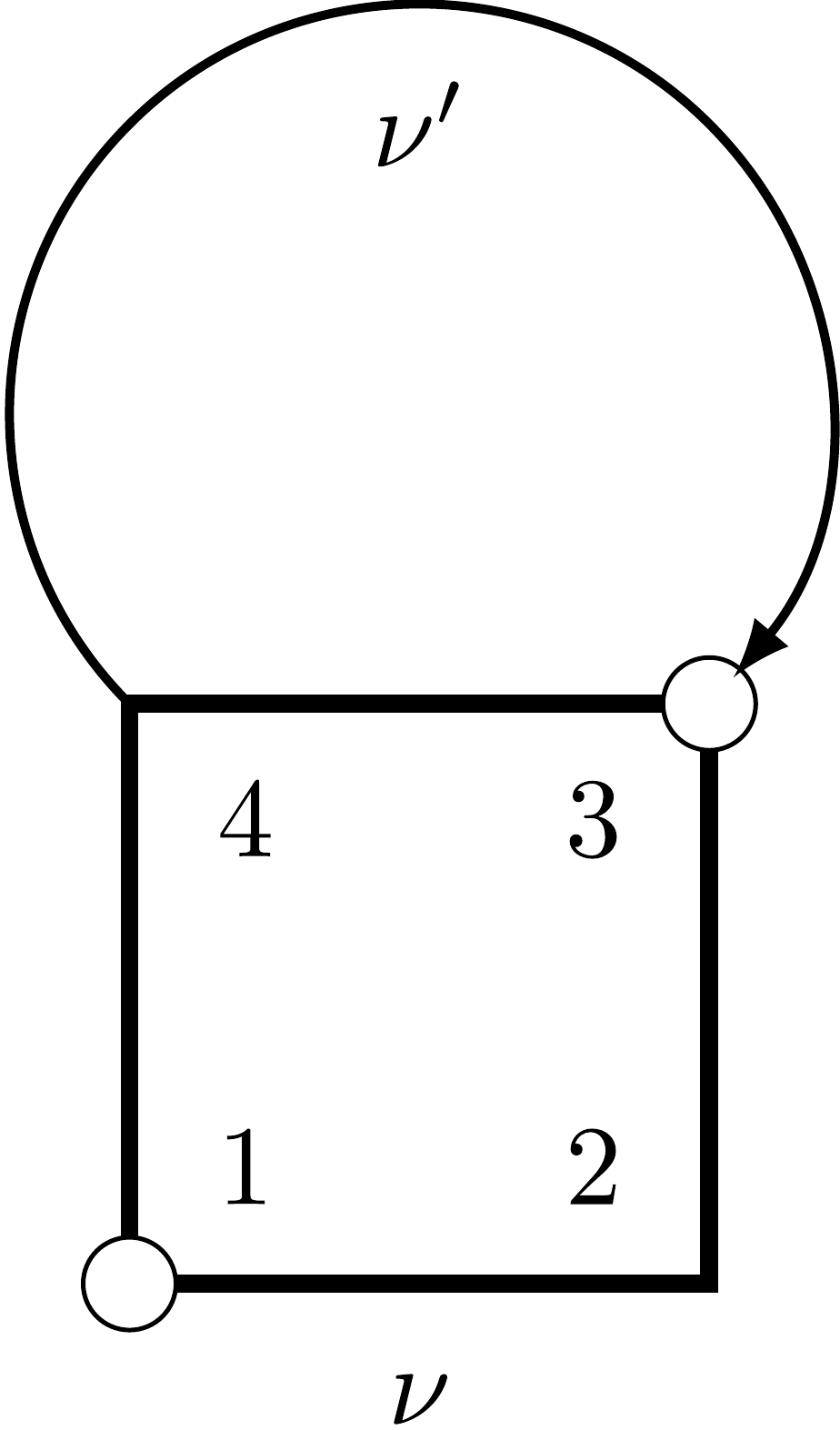}
 \caption{
 Feynman diagram for the first order dual fermion perturbation for the  self-energy $\widetilde{\Sigma}_{12}(\nu)$: a line represents the non-local dual Green's function $\widetilde{G}_{43}(\nu')$ and a box is 
 the two-particle vertex (cumulant) $\gamma_{1234}$, $(\sigma, \sigma')$ are spin-indices.}
\label{fig_1order} 
\end{figure}

The first order for the vertex in particle-hole (PH) channel  
is given by the diagram shown in Fig.{\ref{fig_1order}}

\begin{eqnarray}
\tilde{\Sigma }_{12}^{(1)}=-\sum_{ s-QMC}\sum_{ 3,4  }\gamma_{1234}^{d}(s)\tilde{G}^0_{34}
\label{df:1order}
\end{eqnarray}
Here the density vertex in PH channel is
\begin{eqnarray}
\gamma_{1234}^d=\gamma^{\uparrow \uparrow \uparrow \uparrow  }_{1234}+\gamma^{\uparrow \uparrow \downarrow \downarrow}_{1234}
\label{df:gammad}
\end{eqnarray}
and the final Green's function reads
\begin{equation}
G_{12}=\left[ \left( g+{\tilde{\Sigma}} \right)^{-1}-\tilde{t} \right]^{-1}_{12}
\label{DF_GF}
\end{equation}

Within the determinant DQMC with Ising-fields $\{s\}$ or inside the CT-INT with stochastic sampling of interaction order expancion $\{s\}$ for two-particle correlators we can use the Wick-theorem:

\begin{eqnarray}
 \gamma_{1234}(s)\equiv \langle{c_{1}^{\phantom{*}}c_{2}^{*}c_{3}^{\phantom{*}} c_{4}^{*}} \rangle _s =
 \langle{c_{1}^{\phantom{*}}c_{2}^{*}} \rangle _s \, 
 \langle{c_{3}^{\phantom{*}}c_{4}^{*}} \rangle_s - \langle{c_{1}^{\phantom{*}}c_{4}^{*}}\rangle_s \, 
 \langle{c_{3}^{\phantom{*}}c_{2}^{*}} \rangle _s
\label{gamma4}
\end{eqnarray}

\subsection*{K space scheme}
For large system ($N \ge 4$) it is much faster to calculate the dual self-energy in the K-space with within the QMC Markov chain.
The dual action in K-space reads
\begin{equation}
\tilde{S} [d^{ \ast } ,d] = -\sum _{ \mathbf{k}\, \nu \sigma }d_{ \mathbf{k} \nu \sigma }^{ \ast } \; \tilde{G}_{0\mathbf{k}\nu }^{ -1}\; d_{ \mathbf{k} \nu \sigma } +\frac{1}{4} \sum _{1234  } \gamma_{1 2 3 4} d_{1}^{ \ast } d_{2}^{ \ast } d_{3} d_{4}\, .
\label{eqn::df::Kdual_action}
\end{equation}
Using the short notation $k\equiv (\mathbf{k}, \nu_n)$ and $\nu_n  =(2 n +1) \pi /\beta$, with $n \in \mathbb{Z}$,
the dual Green's function is equal to
\begin{equation}
\tilde{G}_{k}^0  =\left (\tilde{t}_{k}^{ -1} -\hat{g}_{k}\right )^{ -1} \, .
\label{eqn::df::Gdk}
\end{equation}

Since the bare dual Green's function calculated in the independent QMC run for the reference system, it is fully translationally invariant $\tilde{G}^0_{34} \equiv \tilde{G}^0(3-4)$ and we used Fourier transform to calculate the K-space dual Green's function $\tilde{G}^0_k$.

Within the QMC Markov chain the lattice auxilary Green's function is not translationally invariant therefore $g^s_{12}=-\langle c_{1}  c_{2}^{ \ast } \rangle_s $ and we use double Fourier transform to calculate  $g^s_{kk'}$.
To include "disconnected part" of the vertex in equation Eq. (\ref{eqn::chiph}) we just substract exact Green's function from the previus QMC run of the reference system as following
\begin{eqnarray}
\tilde{g}^s_{12}=g^{s}_{12}-g_{12}
\label{gdfqmc}
\end{eqnarray}
In the K-space this subtractions has the following form
\begin{eqnarray}
\tilde{g}^s_{kk'}=g^{s}_{kk'}-g_{k}\delta_{kk'}
\label{gdfqmc}
\end{eqnarray}

For transformation of the vertex $\gamma_{1234}^d$ in Eq. (\ref{df:gammad})within the QMC step in the K-space we take into account that
indices $3,4$ are "diagonal" in $k$-spcae due to multiplication by translationally invariant dual Green's function $\tilde{G}^0_{34}$ which transforms as $\tilde{G}^0_{k}\delta_{kk'}$ and indices $1,2$ become translationally invariant after QMC-summation, which finally leads us
to the following equation for final spin-up components of the first order dual self-energy $\tilde{\Sigma }_{k}$ 
\begin{eqnarray}
\tilde{\Sigma }_{k}^{(1)}=\frac{-1}{(\beta N)^2 Z_{QMC} }\sum_{ s-QMC}\sum_{ k' } \left [  \tilde{g}_{kk}^{\uparrow \uparrow} \tilde{g}_{k' k'}^{\uparrow \uparrow  }    
-  \tilde{g}_{kk'}^{\uparrow \uparrow} \tilde{g}_{k'k}^{\uparrow \uparrow  }  +\tilde{g}_{kk}^{\uparrow \uparrow } \tilde{g}_{k' k'}^{\downarrow \downarrow} \right]_s   \tilde{G}^0_{k'}
\label{1korderQMC}
\end{eqnarray}
Additional normalisation  factor $\frac{1}{(\beta N)^2} $ comes from the Fourier transform in $k$ and from  the $\mathbf{k'}$-sum with $N$-lattice cites and summation over Matsubara frequency: $\frac{1}{\beta} \sum_{\nu'} (...) $. For paramagnetic calculations we average over two spin projections.

Corresponding lattice Green's function reads:
\begin{equation}
G_{k}=\left[ \left( g_k+{\tilde{\Sigma}_k} \right)^{-1}-\tilde{t}_k \right]^{-1} \, .
\label{DF_k}
\end{equation}
Finally, we note, that if we neglect the dual self-energy, $\tilde{\Sigma}_k=0$, this approximation is equivalent to so-called cluster-perturbation theory (CPT) for $N\times N$ system~\cite{Valenti_CPT}.

\begin{center}
{\bf{Supplementary Note 1. Periodization of bare Green's Function}}
\end{center}

For calculation of a bare Green's function for $N\times N$ system we used a special scheme to reduce dependence on
the cluster size. We start from the non-interacting Green's function with given $t'/t$ and $\mu$ for a infinite lattice (in practice $50N\times 50N$ with periodic boundary condition). Then we cut the Green's function to only our small system
$N\times N$ which results in non-periodic Green's function ${\cal G}_{ij}(\nu _n)$ with (i,j=0,N-1). 
In order to periodize the Green's function for small system we average corresponding distance, for example 
${\cal G}_{0,n}$ and ${\cal G}_{0,N/2-n}$. In practice, we used "double" Fourie transform on $i$ and $j$ from ${\cal G}_{ij}$ to ${\cal G}_{kk'}$ and take the diagonal (periodic) part ${\cal G}_{k} \delta_{kk'}$. In this way the local Green's 
function does not depends on the size of our cluster (Supplementary Figure~1a) and the non-local part (Supplementary Figure~1b)  has much faster converge in comparisson to standard periodic DCA cluster scheme~\cite{DCA_RMP}. The reason for this fast 
convergence of the non-local Green's function and exact local Green's function related with real space periodization,
while DCA make the average patches in the K-space.

\begin{figure}[t!]
\includegraphics[width=.45\linewidth]{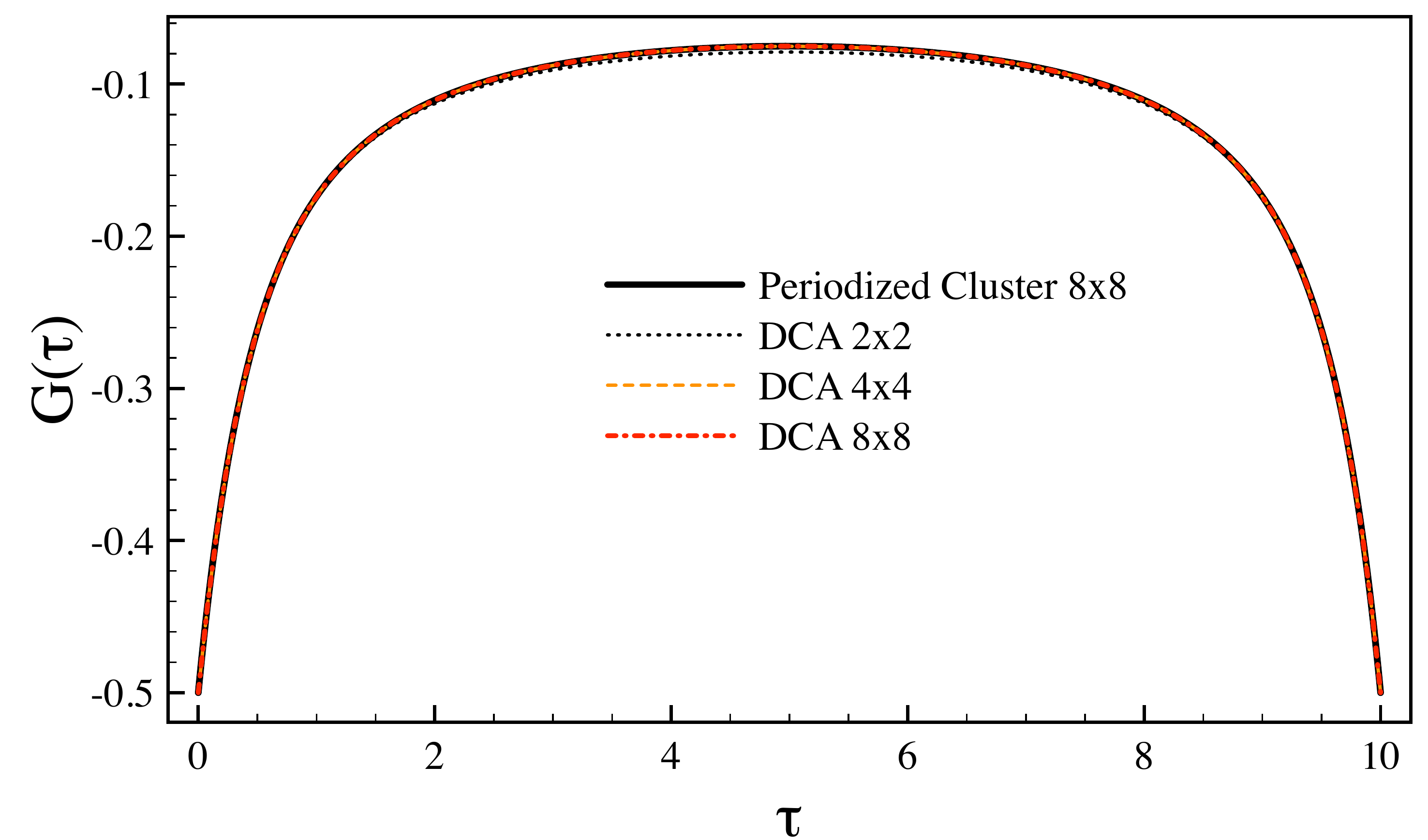}
\includegraphics[width=.45\linewidth]{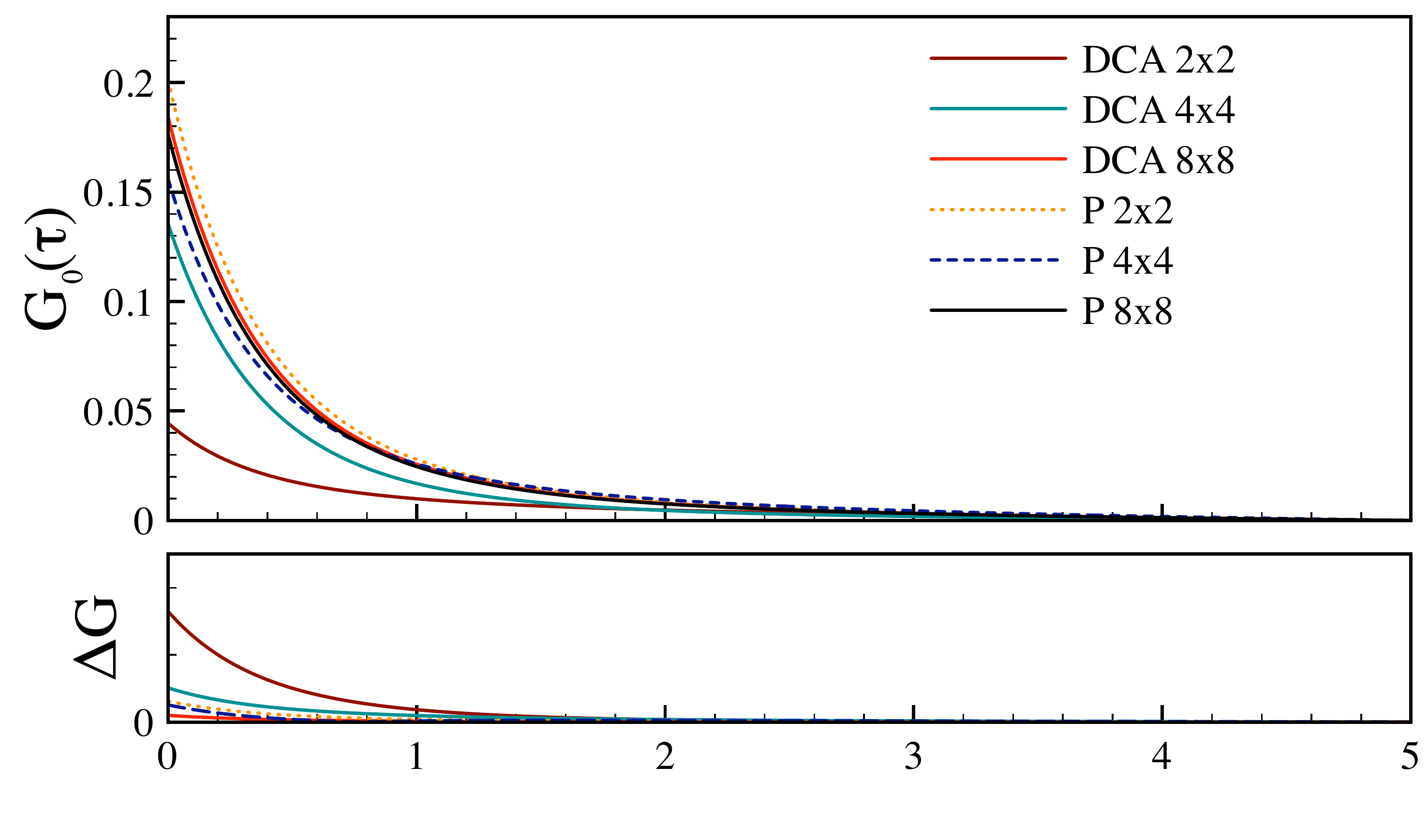}
\begin{center}
{Supplementary Figure 1.  
Periodized Green's function in imaginary time $\tau$ for clusters $N\times N$ with $N=2,4,8$ for our scheme compare 
to DCA approach for the reference system with $U=5.56$ $t'=0$ and $\mu=0$ for $\beta=10$. The local Green's function (left) and  first nearest-neighbours (right).
}
\end{center}
\label{fig:Xi1221}
\end{figure}

\begin{center}
{\bf{Supplementary Note 2.  Tests for the $2 \times 2$ system}}
\end{center}

For the smallest $2\times 2$ we can use a very transparent real-space DF-QMC algorithm.
In this case we do not need any additional  periodization since  $2\times 2$ cluster is ``self-periodic''.
Since there is almost no sign problem in DQMC method for the doped $2\times 2$ cluster in the bath,
we can compare the first-order dual-fermion perturbation with numerical exact DQMC results.
The all three non-equivalent Green's functions for $2\times 2$  system are shown in the
Supplementary Figure~2 using first-order DF-correction within Hirsch-Fye QMC formalism. 
For small perturbation  $\Delta \mu=-0.3$ and $\Delta t'=0$ a comparison 
with exact DQMC results (point on Supplementary Figure~2) is perfect.
For a large perturbation  $\Delta \mu=-1.5$ and $\Delta t'=0.15$ one can already see a small difference 
from the exact DQMC Green's function. Nevertheless, the results of DF-QMC with only first-order
corrections for the dual self-energy are very satisfactory. Note that for square lattice the
perturbation in $\Delta \mu$ one can compare with $4\Delta t'$, also dispersion of $t'$ terms
made this estimation questionable.

\begin{figure}[t!]
\includegraphics[width=.45\linewidth]{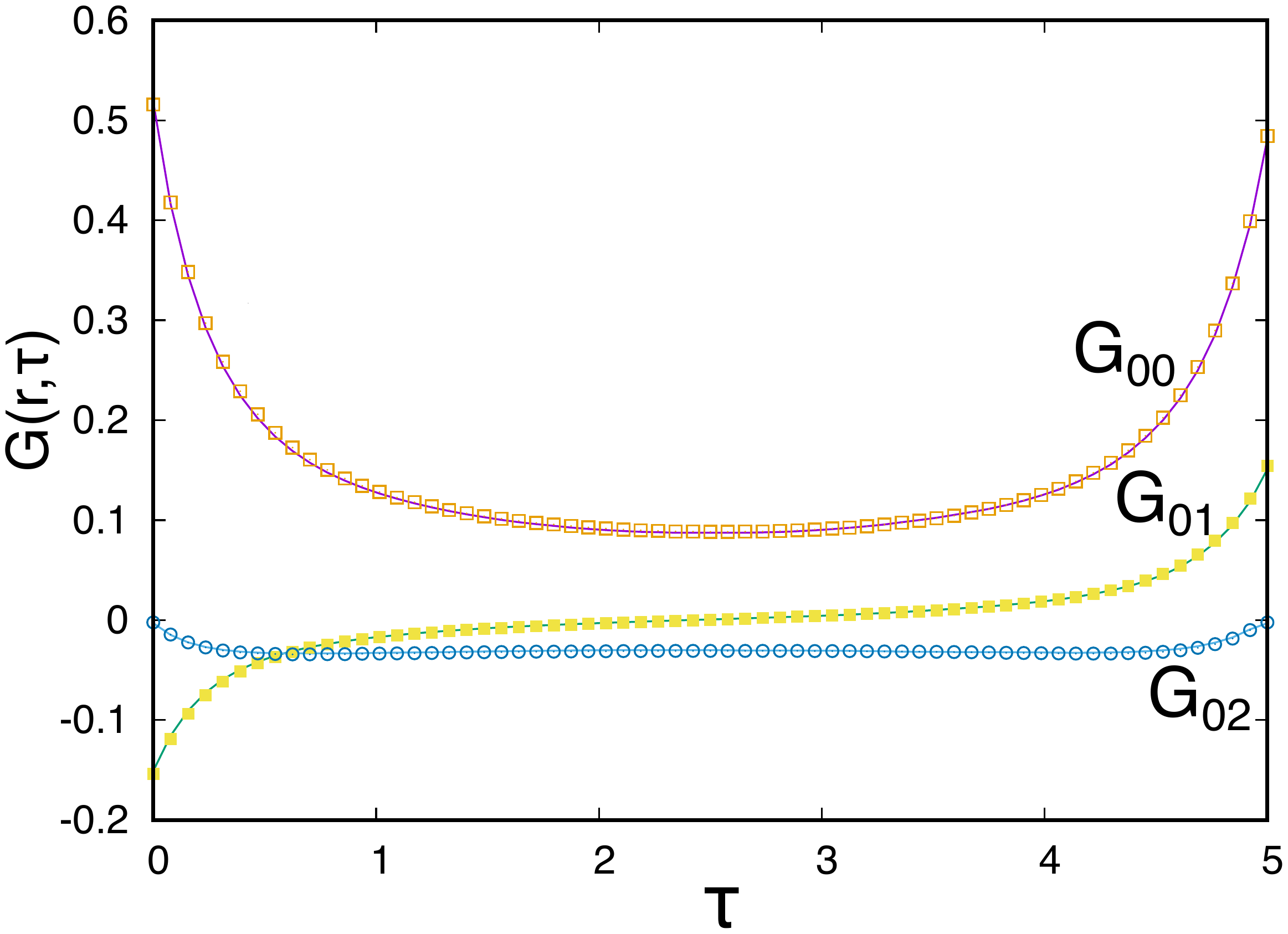}
\includegraphics[width=.45\linewidth]{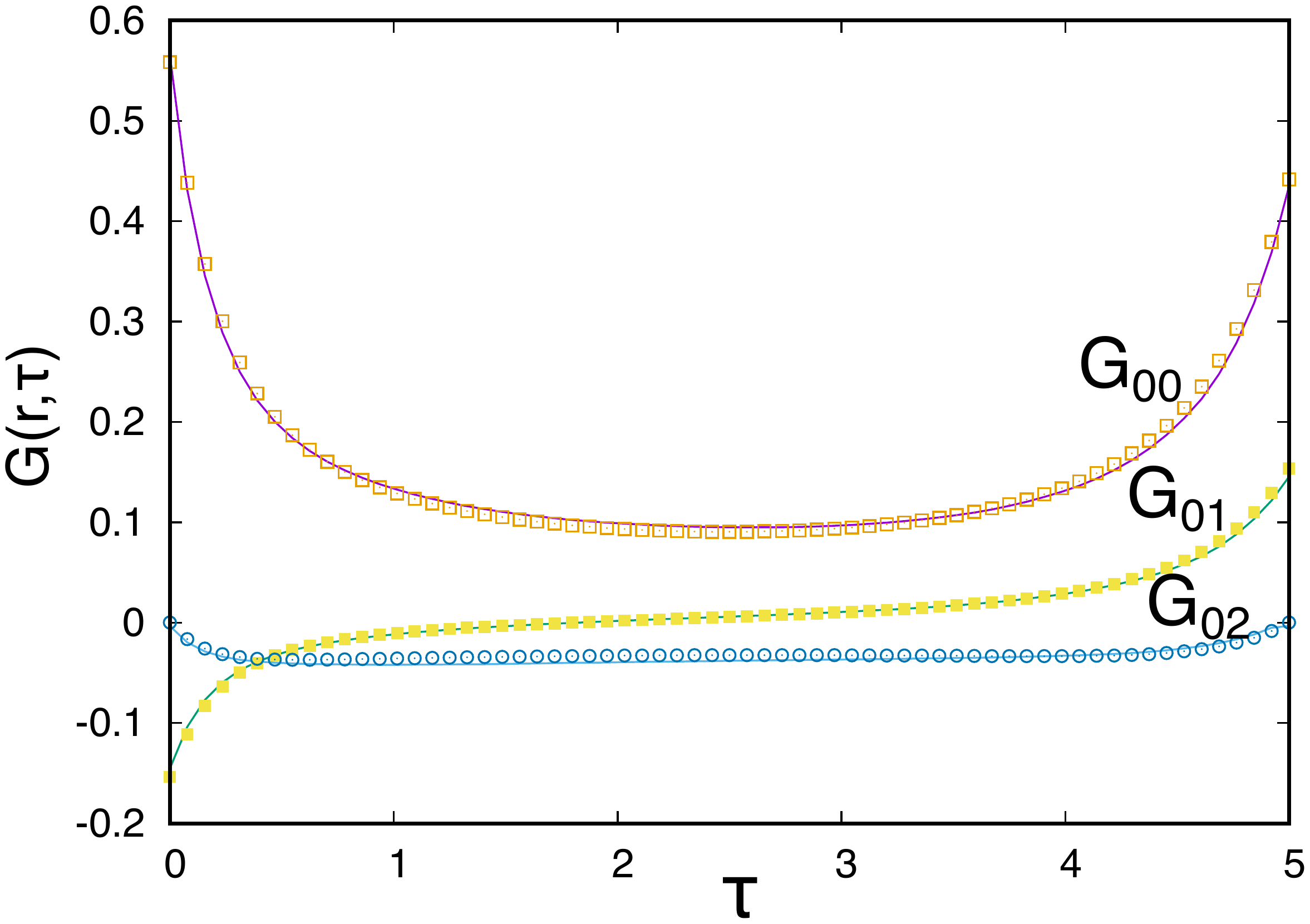}
\begin{center}
{Supplementary Figure 2.  
Three non-equivalent components of the  Green's functions for  $2\times 2$ system as function of imaginary time 
for $U=5.56$, $\beta=5$ and  $t'/t=-0.1$, $\mu=0$ (left) and  $t'/t=-0.3$, $\mu=-1.3$ (right). Notes, that here we use a QMC defitition with positive local Green's function.}
\end{center}
\label{fig:Xi1221}
\end{figure}

\begin{center}
{\bf{Supplementary Note 3.:  Tests for the $4 \times 4$ system}}
\end{center}

\begin{figure}[t!]
\includegraphics[width=.49\linewidth]{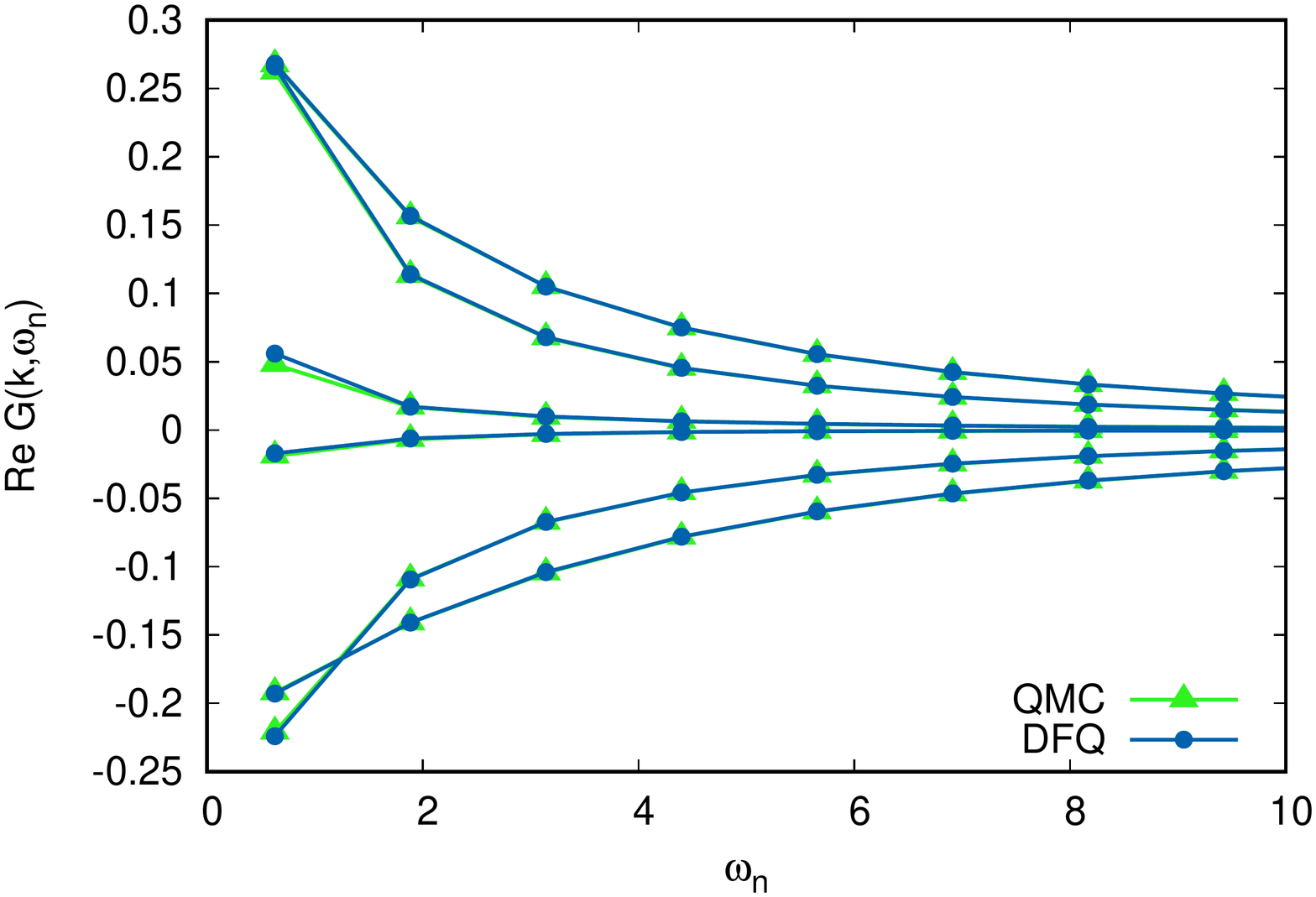}
\includegraphics[width=.49\linewidth]{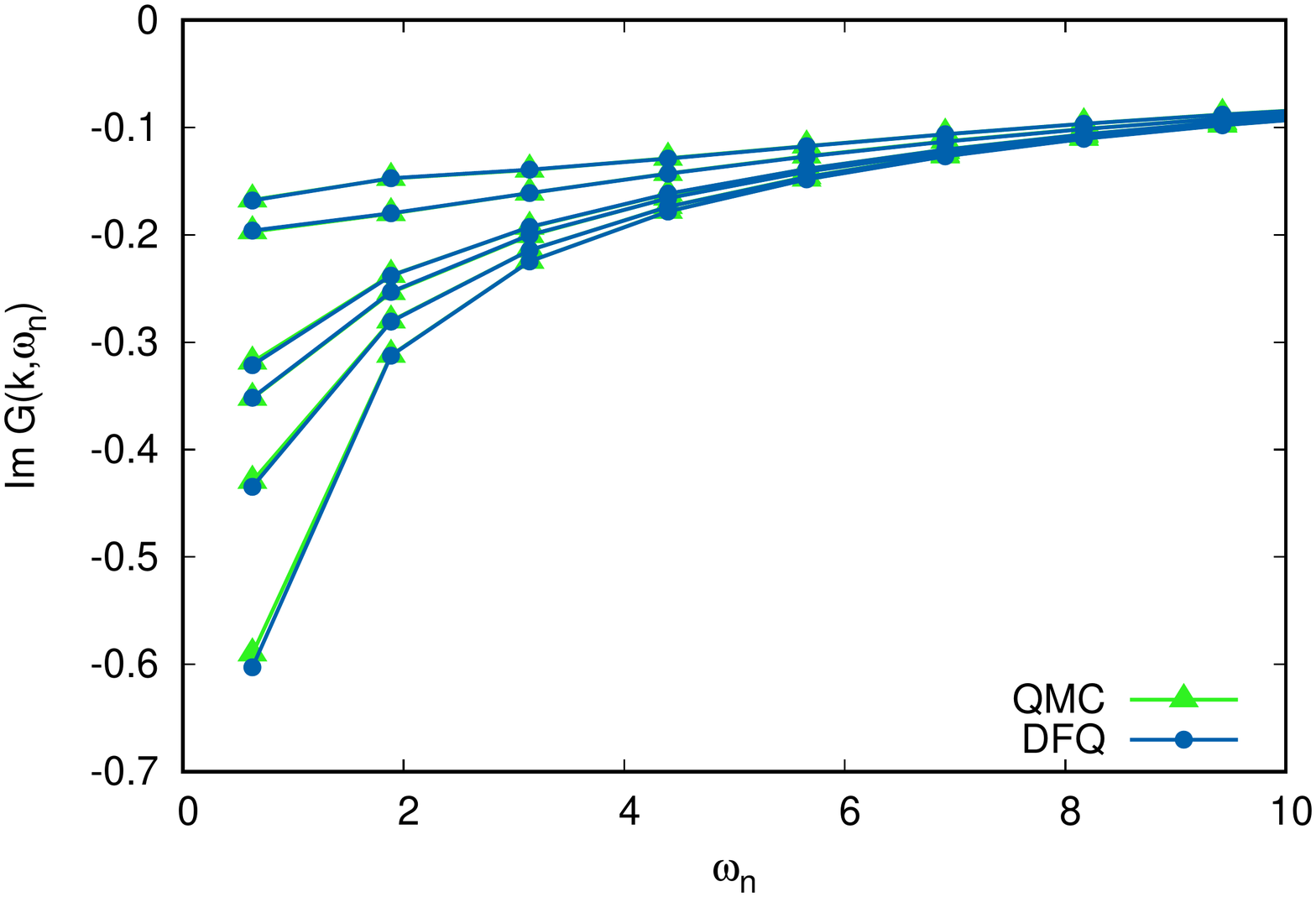}
\begin{center}
{Supplementary Figure 3.  
 Green's functions from DF-QMC (DFQ) in comparison with numerically exact DQMC result (QMC) for the (4 $\times $ 4) system in Matsubara space with $U=5.56$, $t'/t=-0.1$, $\mu=0$  and $\beta=5$, real par (left) and imaginary part (right).}
\end{center}
\label{fig:EgED7x7}
\end{figure}
\begin{figure}[t!]
\includegraphics[width=.49\linewidth]{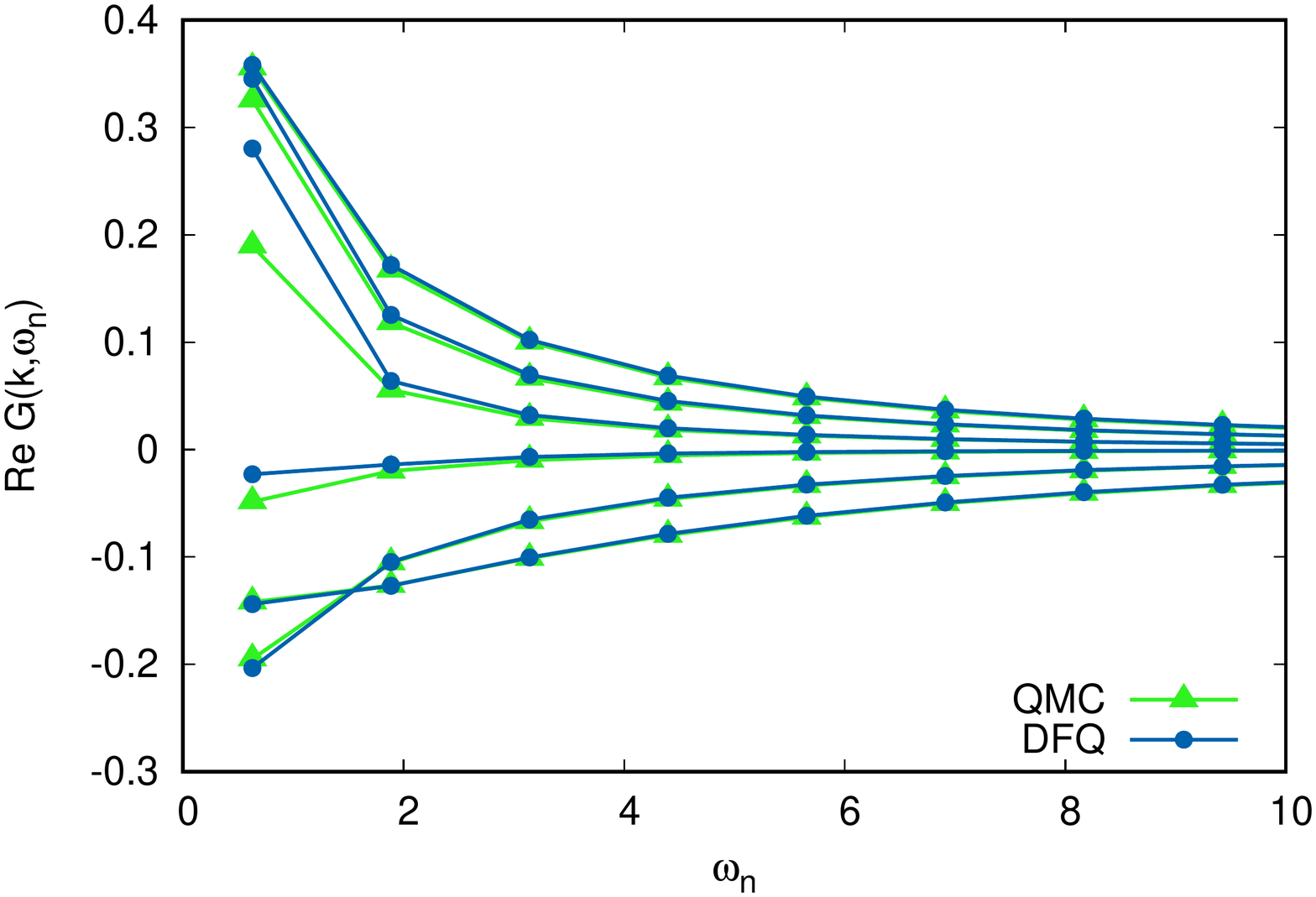}
\includegraphics[width=.49\linewidth]{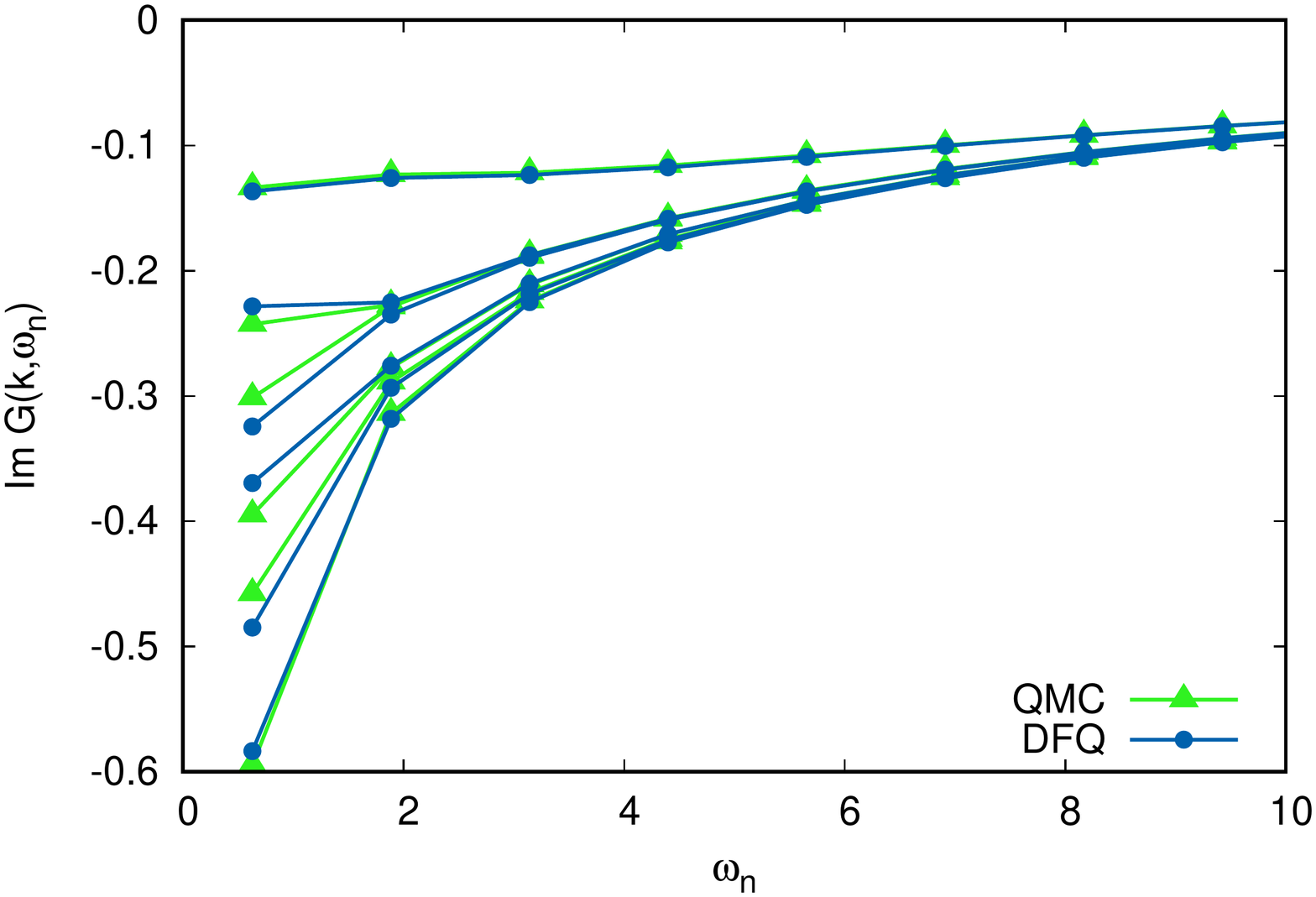}
\begin{center}
{Supplementary Figure 4.  
 Green's functions from DF-QMC (DFQ) in comparison with numerically exact DQMC result (QMC) for the (4 $\times $ 4) system in Matsubara space with $U=5.56$, $t'/t=-0.3$, $\mu=0$  and $\beta=5$, real par (left) and imaginary part (right).}
\end{center}
\label{fig:EgED7x7}
\end{figure}
\begin{figure}[t!]
\includegraphics[width=.49\linewidth]{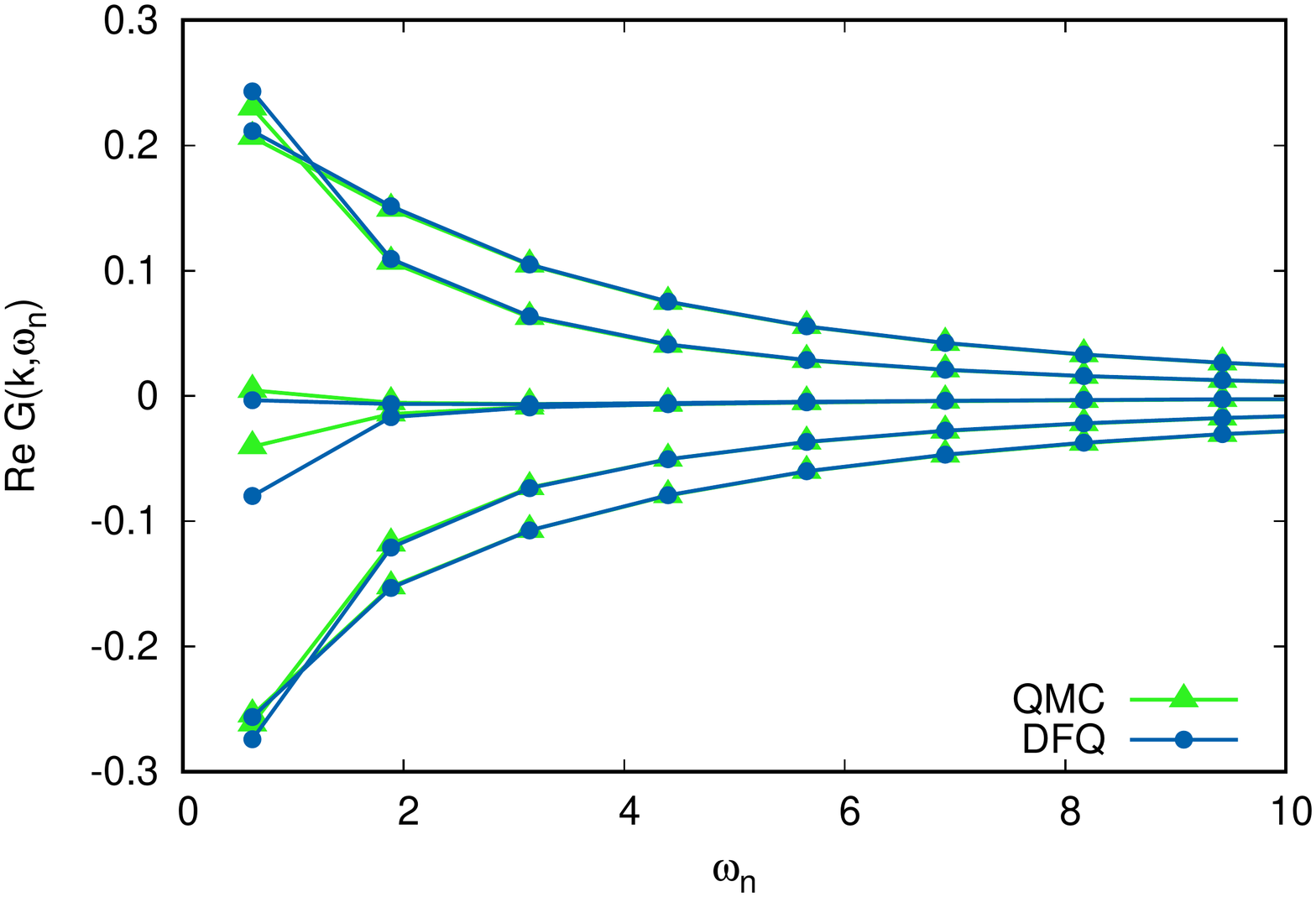}
\includegraphics[width=.49\linewidth]{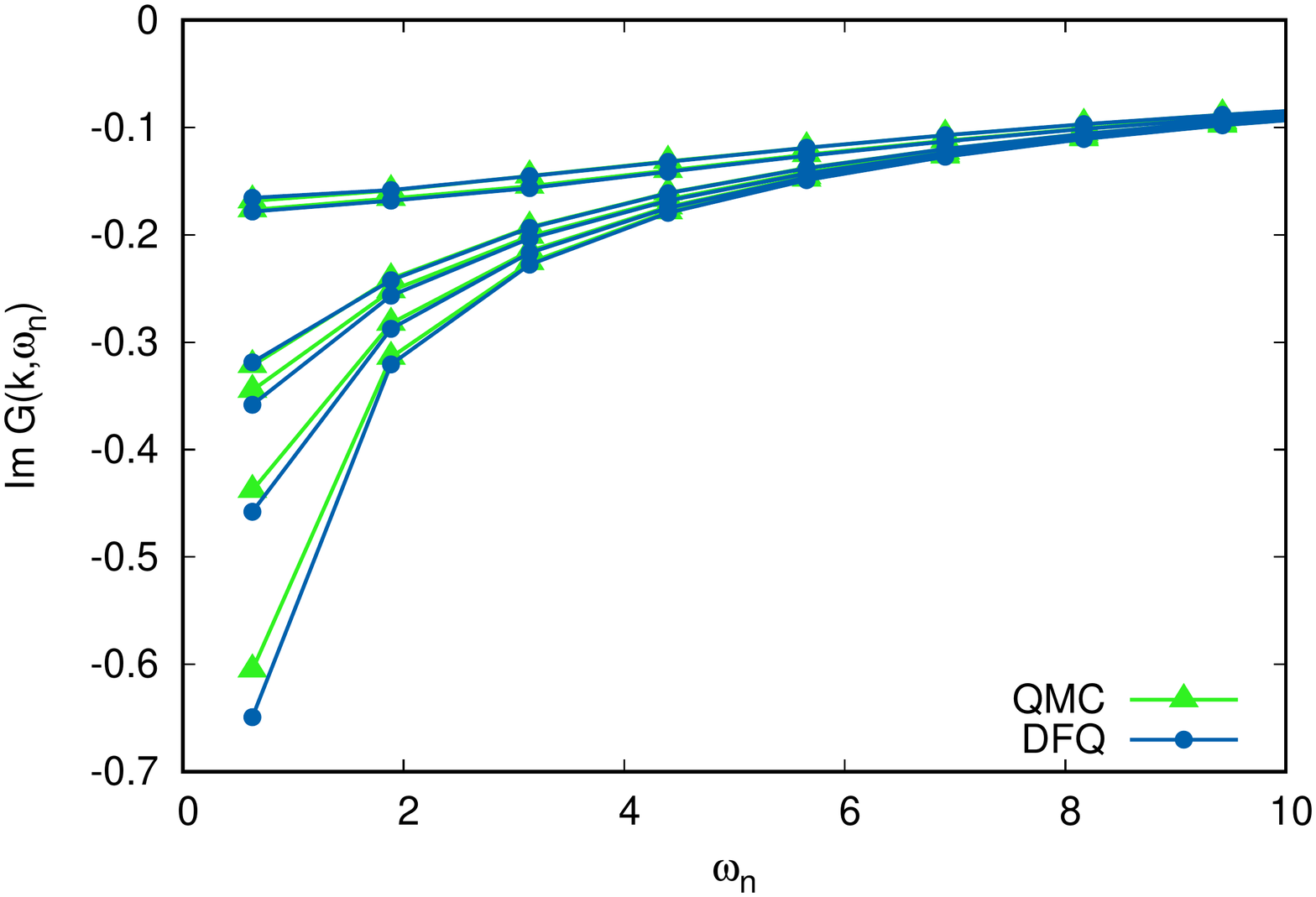}
\begin{center}
{Supplementary Figure 5.  
 Green's functions from DF-QMC (DFQ) in comparison with numerically exact DQMC result (QMC) for the (4 $\times $ 4) system in Matsubara space with $U=5.56$, $t'/t=0$, $\mu=-0.5$  and $\beta=5$, real par (left) and imaginary part (right).}
\end{center}
\label{fig:EgED7x7}
\end{figure}

We analyse performance of DF-QMC formalism as function of  $\Delta \mu$  and $\Delta t'$ for the 4 $\times $ 4 periodic cluster with and without fernionic bath. The DQMC sign problem for 4 $\times $ 4 system is also mild and we can
compare our DF-QMC with numerically exact test for the same $\mu$  and $t'$.
We use value of $U=5.56$ which corresponds to the degenerate ground state of plaquete \cite{danilov2022}.
For all Hirsch-Fye DQMC calculations we use imaginary time discretisation with $L=64$ slices.
Supplementary Figure~3 shows DF-QMC results for a small perturbation $t'/t=-0.1$ and $\beta=5$ in comparison
with exact DQMC results. The agreement is very good which show the strength dual fermion QMC theory.
for a small perturbation. Next, we compare Supplementary Figure~4  for the $t'/t=-0.3$ case which corresponds to optimal  next-nearest hopping in  cuprate materials. In this case one can see the difference from exact DQMC results on the first
Matsubara frequency, but still the overall agreement in all 6 nonequilibrium k-points for 4 $\times $ 4 system
is quite satisfactory.

The effects of chemical potential shift $\mu=-0.5$ presented on Supplementary Figure~5. Qualitatively, agreement
between perturbative DF-QMC  and exact DQMC is similar to the case of $t'/t=-0.3$, but the structure of all
6 Green's functions for  $4\times 4$  system is very different and still the dual-fermion strong coupling 
perturbation in K-space works reasonably well.  

Finally, Supplementary Figure~6, shows the combine effects of
strong chemical potential shift $\mu=-1$ and next-nearest hopping $t'/t=-0.3$. 
In this case we shows results of the CT-INT approach with $\beta=10$
The agreement is very good, and only one Green's function (the lowest  imaginary part) which corresponds to $X$-point ${\bf k}=(\pi,\pi)$ and located close to the Fermi level (corresponding real part is close to zero) the dual perturbation shows small discrepancy.  
In principle, one can  reduce the error of dual perturbation if one can choose the reference more close to the target system. The main condition is a weak
sign problem for the reference system. Since it is always the case for small
$4 \times 4$ cluster, we show in the Supplementary Figure~7 the CT-INT results of reference system corrrsponding to $t'/t=-0.1$ for the target system of 
$t'/t=-0.3$. In this case, for lower temperature  $T=t/10$ the DF-QMC 
resalts are still in a good agreement with exact solution.

\begin{figure}[t!]
\includegraphics[width=.47\linewidth]{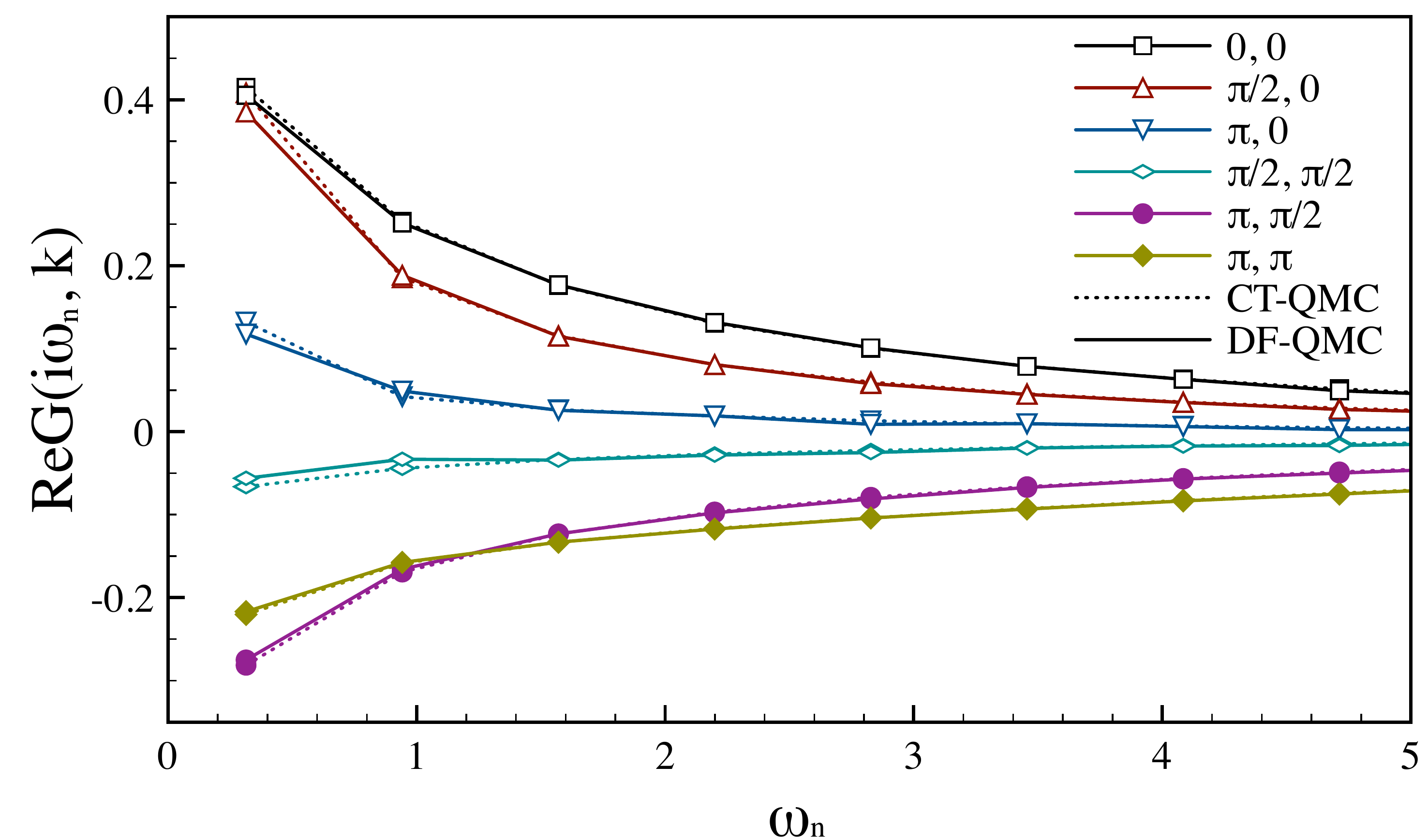}
\includegraphics[width=.47\linewidth]{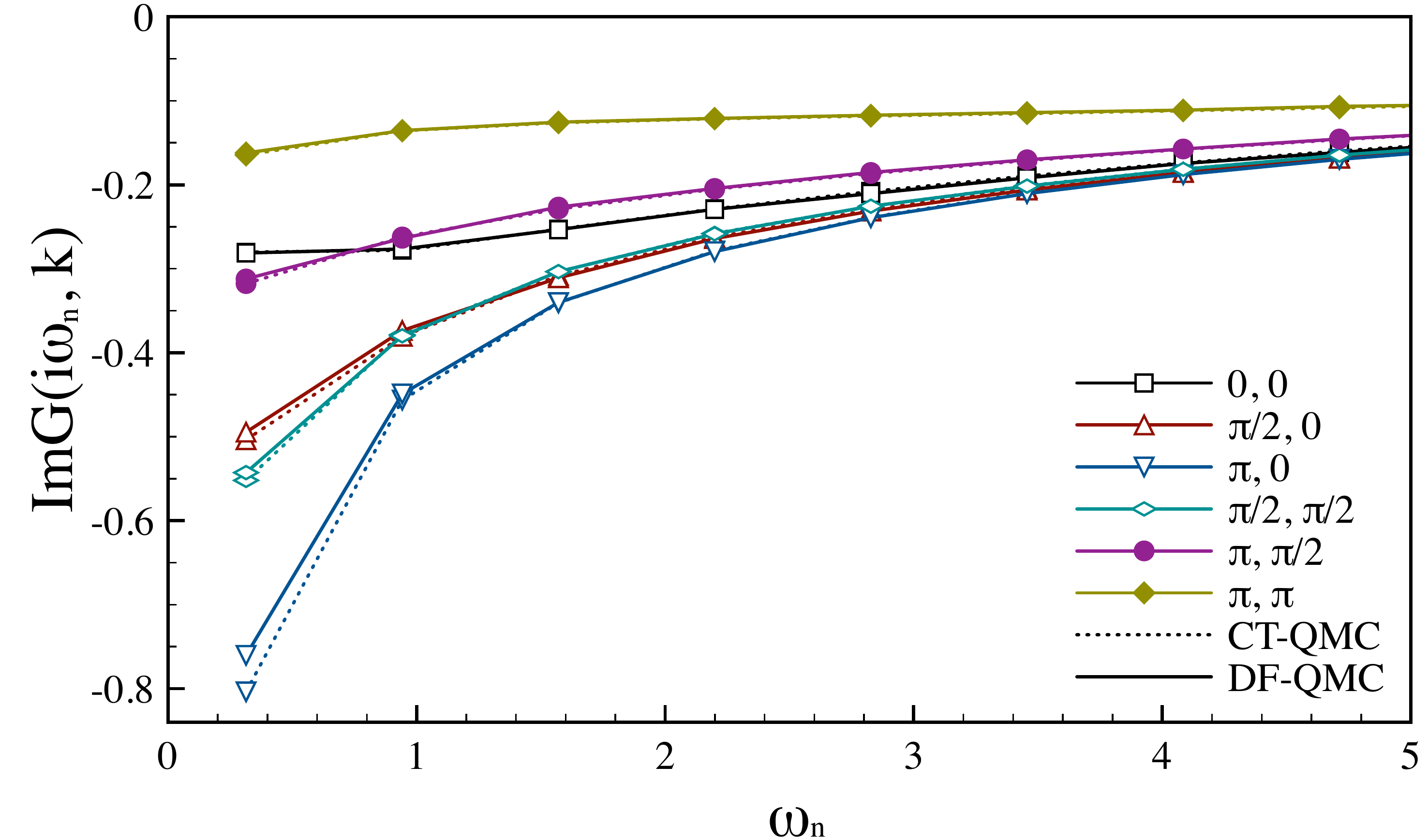}
\begin{center}
{Supplementary Figure 6.  
 Green's functions from DF-CT-QMC (DF-QMC) in comparison with numerically exact QMC result (CT-QMC) for the (4 $\times $ 4) system in Matsubara space with $U=5.56$, $t'/t=-0.3$, $\mu=-1$  and $\beta=10$, real par (left) and imaginary part (right).}
\end{center}
\label{fig:EgED7x7}
\end{figure}
\begin{figure}[t!]
\includegraphics[width=.47\linewidth]{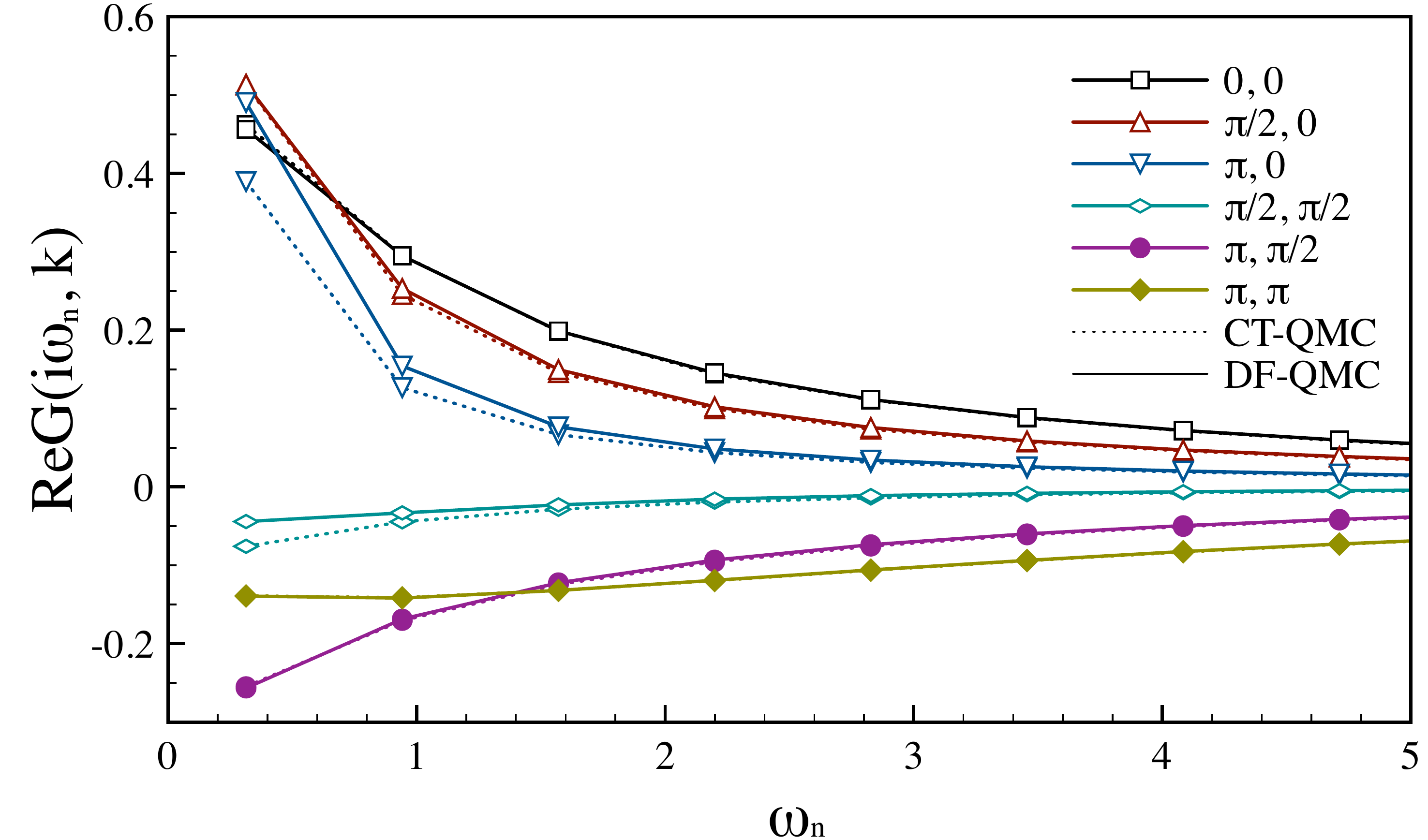}
\includegraphics[width=.47\linewidth]{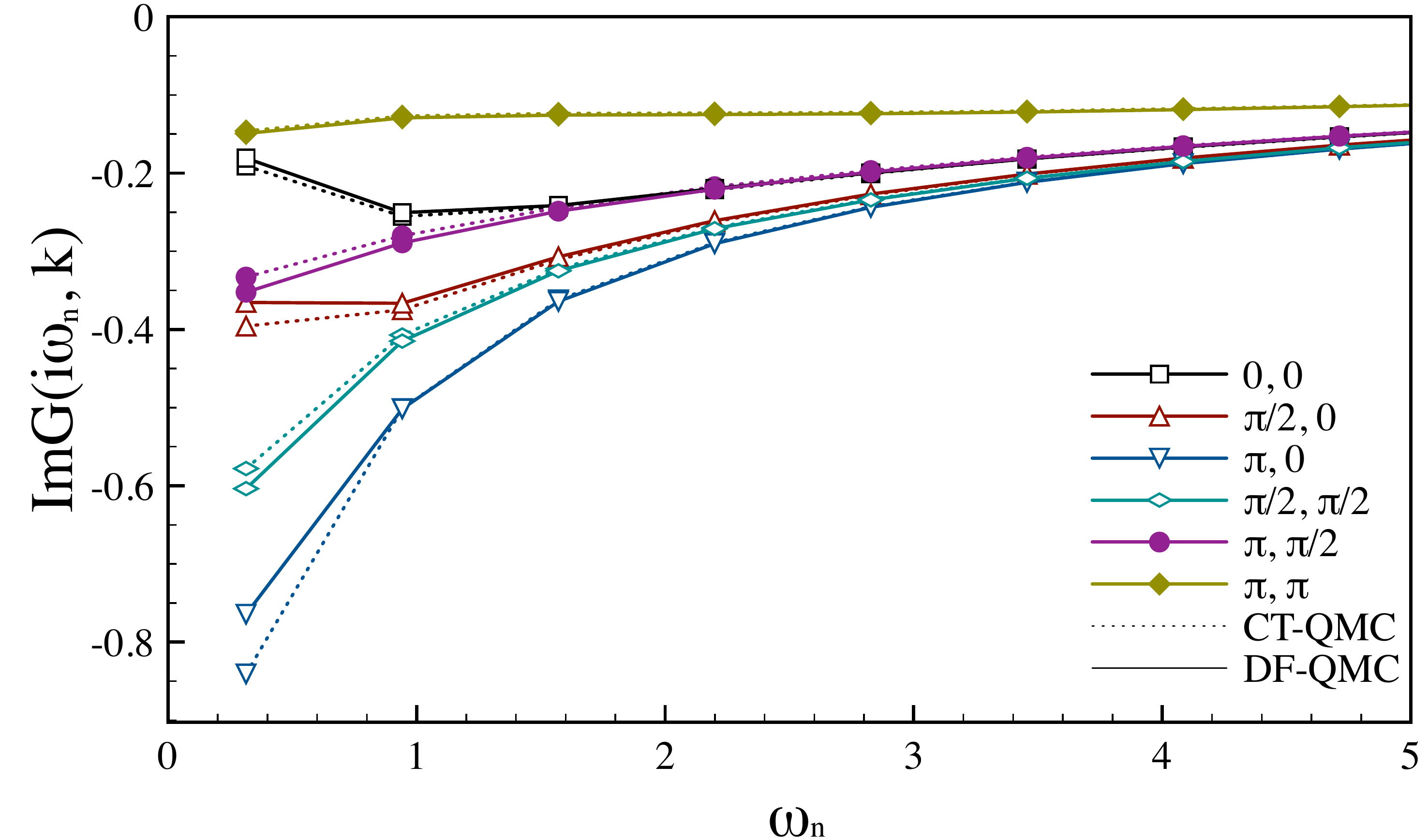}
\begin{center}
{Supplementary Figure 7. 
 Green's functions from DF-CT-QMC (DF-QMC) perturbation for $t'/t=-0.3$, starting from the reference system with $t'/t=-0.1$ in comparison with numerically exact QMC result (CT-QMC) for the (4 $\times $ 4) system in Matsubara space with $U=5.56$,  $\mu=0$  and $\beta=10$, real par (left) and imaginary part (right).}
\end{center}
\label{fig:DOS4x4}
\end{figure}

\begin{figure}[t!]
\includegraphics[width=.49\linewidth]{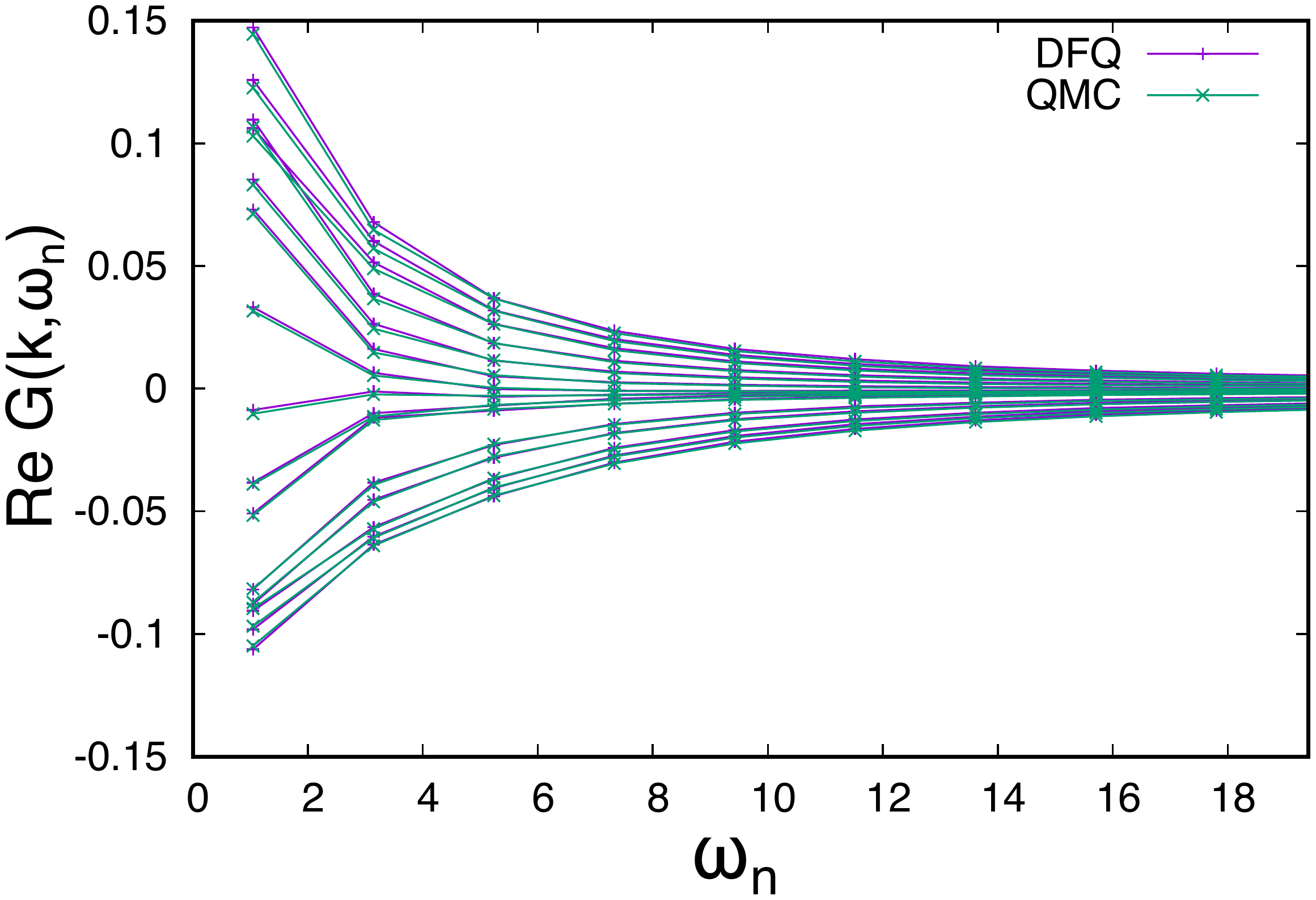}
\includegraphics[width=.49\linewidth]{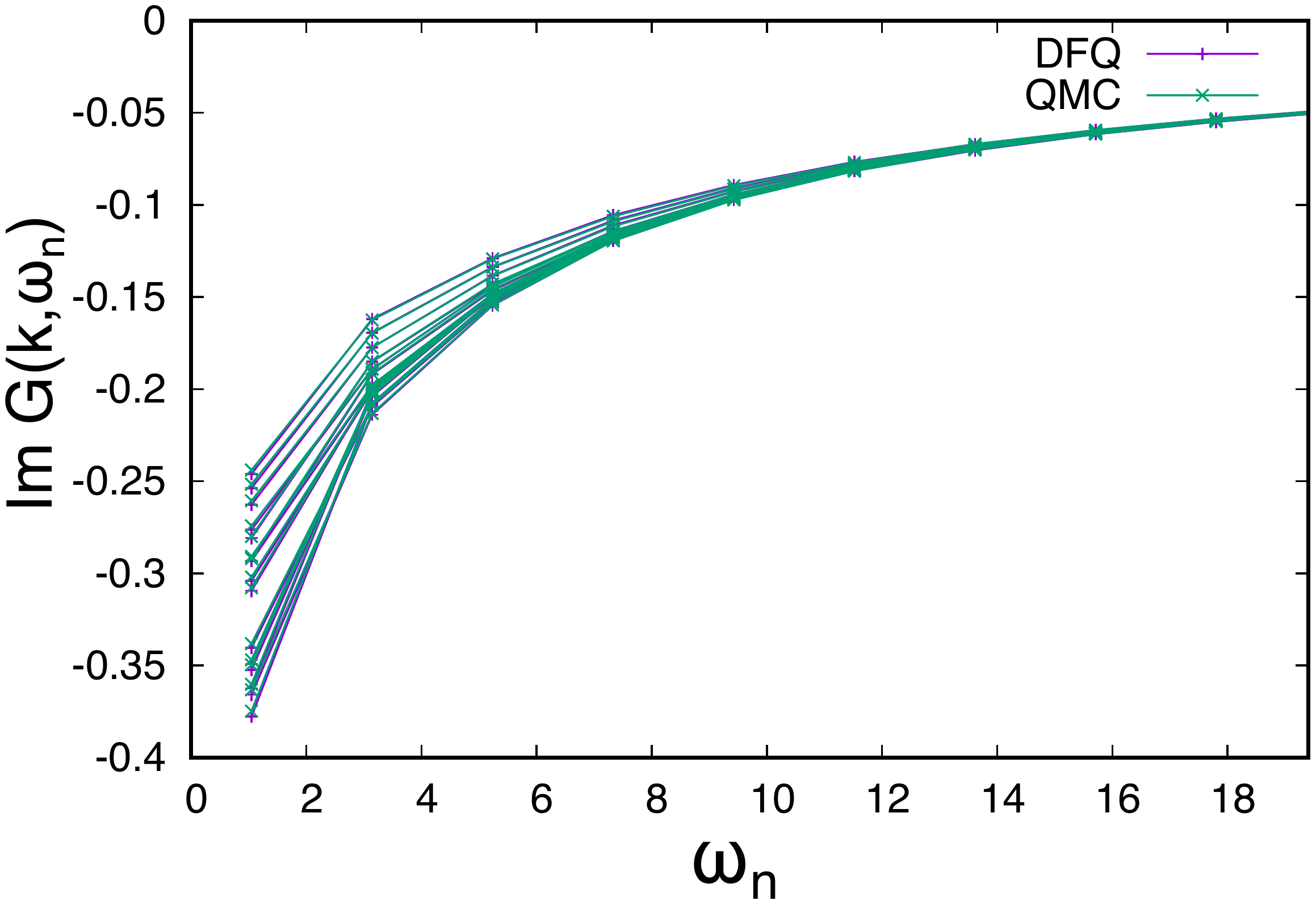}
\begin{center}
{Supplementary Figure 8. 
Green's function for the (8 $\times $ 8)  for $U=5.56$, $t'/t=-0.3$, $\mu=-1.5$ $\beta=3$ for Hirsch-Fye DF-QMC (DFQ) in
comparisson with DQMC test (QMC).
}
\end{center}
\label{fig:DOS4x4}
\end{figure}

\begin{figure}[t!]
\includegraphics[width=.44\linewidth]{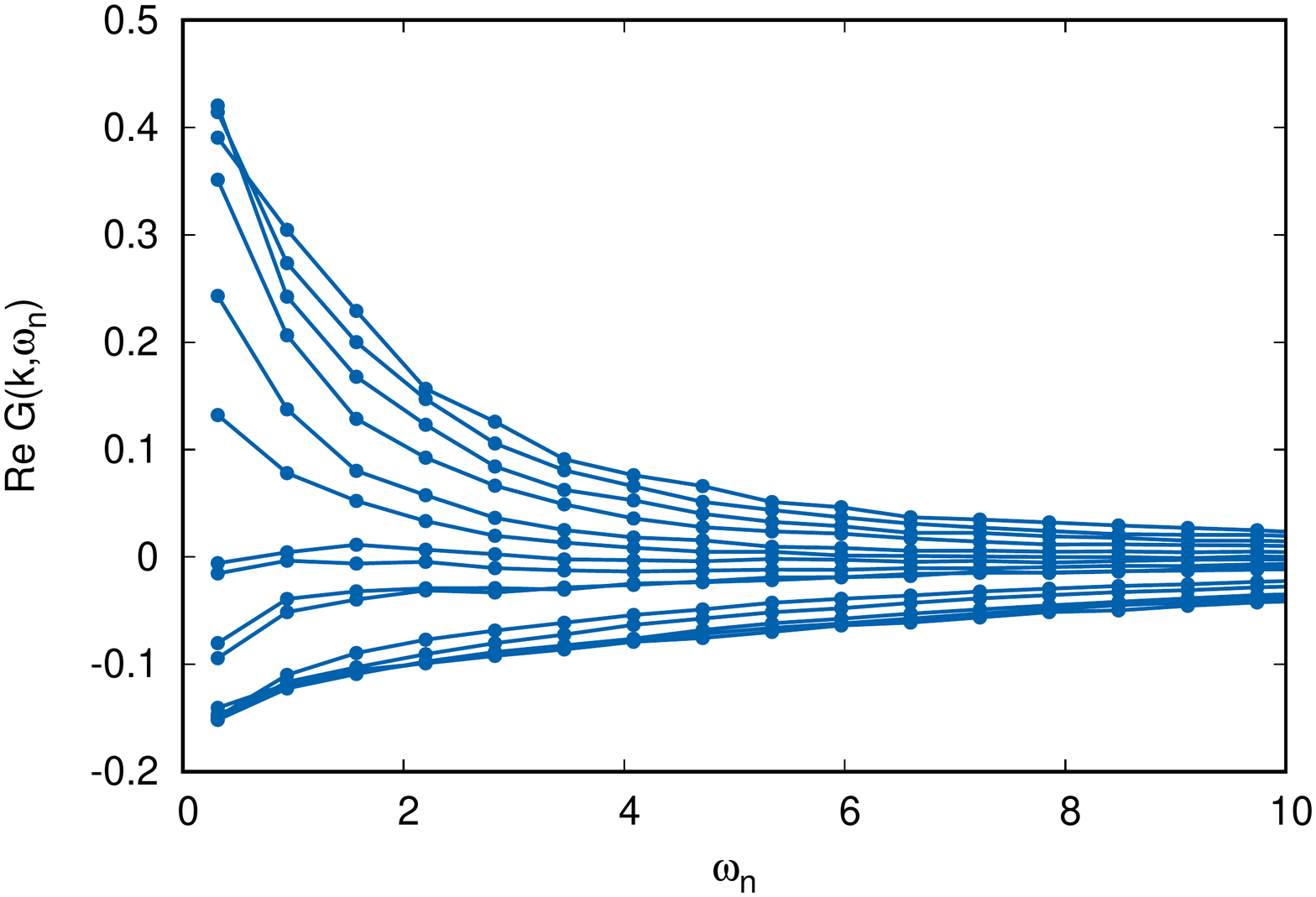}
\includegraphics[width=.44\linewidth]{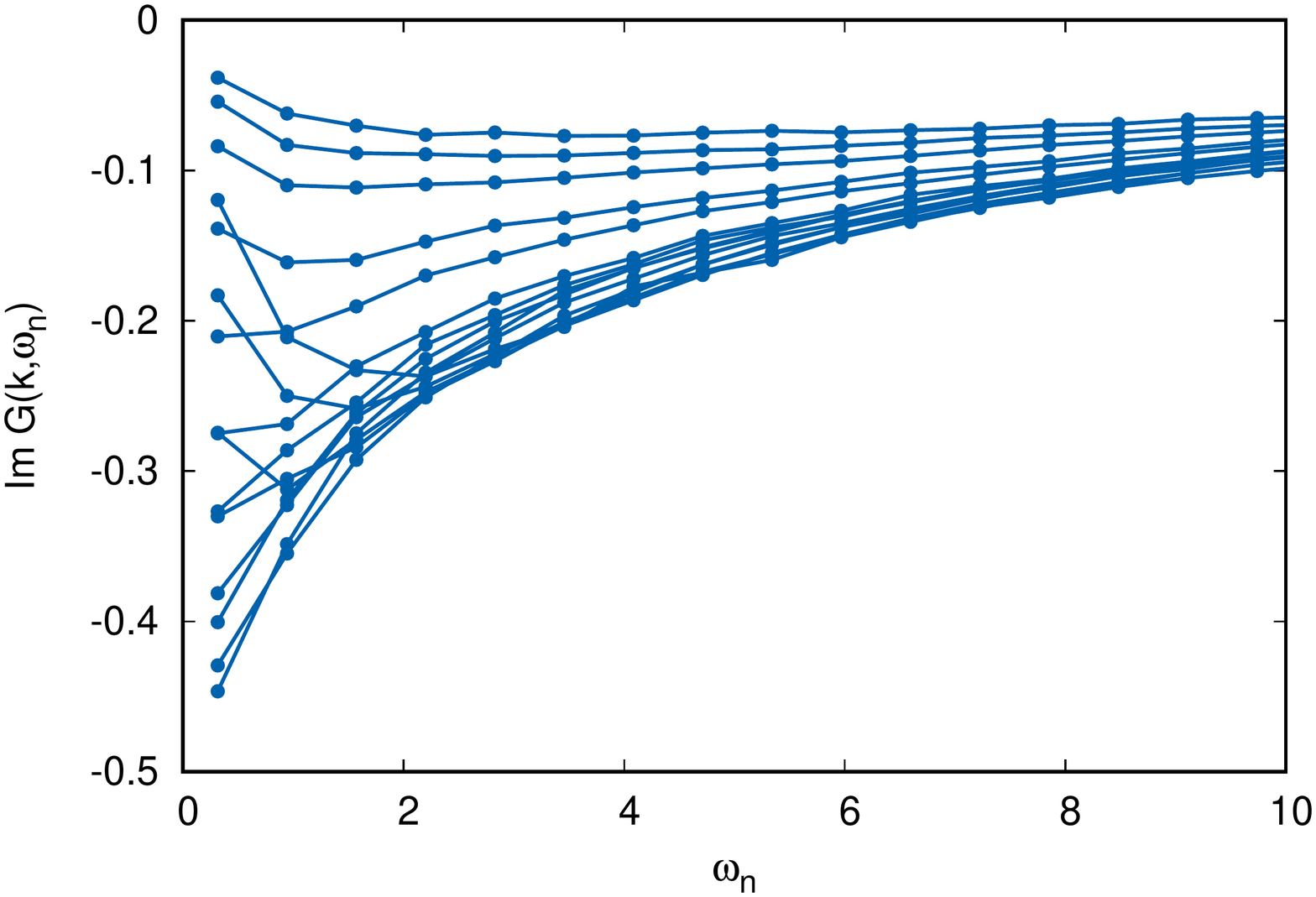}
\begin{center}
{Supplementary Figure 9. 
Green's functions within Hirsch-Fye DF-QMC scheme for 15 nonequivalent $\bf k$-points of the (8 $\times $ 8) cluster in the bath with $U=8$, $\mu=-2$, $t'/t=-0.3$ and $\beta=10$
}
\end{center}
\label{fig:DOS4x4}
\end{figure}

\begin{figure}[t!]
\includegraphics[width=0.45\linewidth]{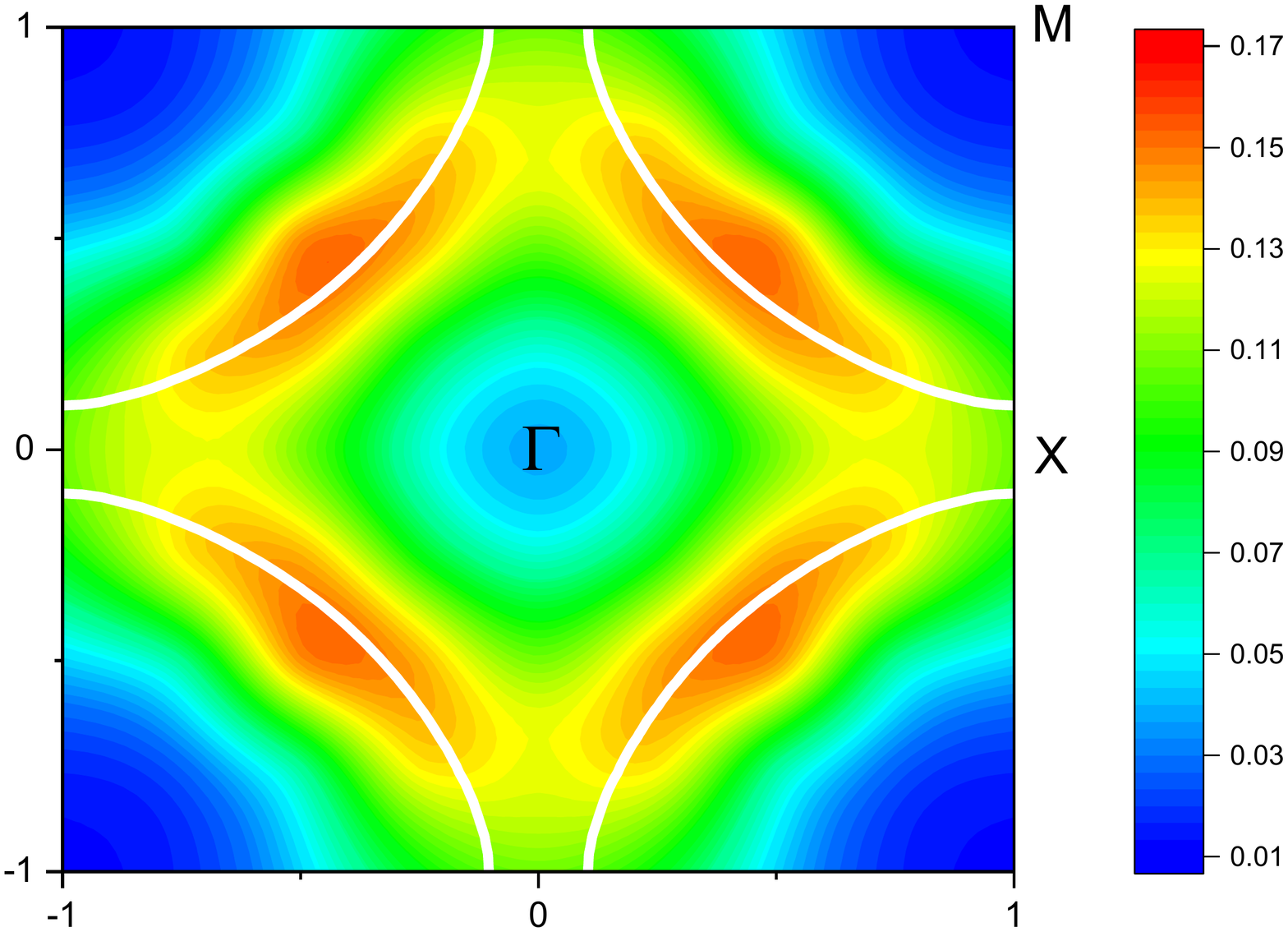}
\begin{center}
{Supplementary Figure 10.  
The Fermi surface calculating within Hirsch-Fye DF-QMC scheme forthe (8 $\times $ 8) cluster in the bath with $U=8$, $\mu=-2$, $t'/t=-0.3$ and $\beta=10$.} 
\end{center}
\label{fig:cdicj}
\end{figure}

\begin{center}
{\bf{Supplementary Note 4.:  Tests for the $8 \times 8$ system}}
\end{center}

For large $8 \times 8$ system for $\beta=10$ the average sign in DQMC is of the order of a $10^{-3}$ even with a fermionic bath and calculation of the test Green's function is not any more possible. 
For much larger temperature, corresponds to $\beta=3$ and not so large $U=5.5.6$ the sign problem is not
severe and it is still possible to prepare a DQMC test. Supplementary Figure~8 shows a comparison of
Matsubara Green's function for all 15 non-equivalent $\bf k$-points im dual-fermion perturbation 
with Hirsch-Fye QMC code and DQMC-test.
The agreement is quite good, but one should remember a very high temperature of this test ($T=t/3$) which
results in metallic behaviour of all Green's functions.
We also present the Green's functions for the same parameters as in the main text, but using the Hirsch-Fye QMC
schem for  $8 \times 8$ system (Supplementary Figure~9) and corresponding Fermi surface (Supplementary Figure~10).
There is a good agreement with CT-INT scheme in the main text, and if we take into account very different
computational scheme for these two methods, we can state that "as proof of principle" the DF-QMC approach
can efficiently overcome the sign problem in the lattice DQMC approach for doped systems with long-range
hoppings.

\vskip 0.5cm

The authors thank Alexei Rubtsov, Evgeny Stepanov, Igor Krivenko, J{\"o}rg Schmalian, Richard Scalettar, Emanuel Gull, Fedor \v{S}imkovic IV, Riccardo Rossi
and Antoine Georges for valuable comments on the work. 

This work was partially supported by the Cluster of Excellence ``Advanced Imaging of Matter'' of the Deutsche Forschungsgemeinschaft (DFG) - EXC 2056 - Project No. ID390715994,  by European Research Council via Synergy Grant 854843 - FASTCORR.
The part of simulations were performed on the national supercomputer HPE Apollo Hawk at the High Performance Computing Center Stuttgart (HLRS) under the grant number QMCdynCOR/44167. User Facility supported by the Office of Science of the U.S. Department of Energy under Contract No. DE-AC02-05CH11231 using NERSC award BES-ERCAP0020359.

\bibliography{Ref}

\clearpage

\end{document}